\documentclass[pre,floatfix,superscriptaddress,preprint,showpacs]{revtex4}
\usepackage{latexsym}
\usepackage{amstext,amsfonts,amssymb,amsmath}
\usepackage{rotating}
\def\sigmaconf{\{\boldsymbol{\sigma}\}_{\rm conf}}

\def\sigmabold{\boldsymbol{\sigma}}

\def\TboldNr{\boldsymbol{\rm T}(N,r)}

\def\TboldNri{\boldsymbol{\rm T}_i(N,r)}

\def\F0{F^{(0)}(\ell_2,\ell_1)}

\def\vxi{\vec{\xi}}
\def\vxiconf{\{\vec{\xi}(\ell_2,\ell_1,\ell_0)\}_{\rm conf}}
\def\vxiconfshort{\{\vec{\xi}\}_{\rm conf}}
\usepackage{epsfig}
\begin{document}
\title{The distribution of the number of node neighbors in random hypergraphs}
\author{Eduardo L\'{o}pez}
\affiliation{CABDyN Complexity Centre, Sa\"{\i}d Business School, University of Oxford, Park End Street,
Oxford OX1 1HP, United Kingdom}
\affiliation{Physics Department, Clarendon Laboratory, University of Oxford, Parks Road, Oxford OX1 3PU,
United Kingdom}
\email{eduardo.lopez@sbs.ox.ac.uk}
\date{\today}
\begin{abstract}
Hypergraphs, the generalization of graphs in which edges become conglomerates of $r$ nodes called hyperedges
of rank $r\geq 2$, are excellent models to study systems with interactions that are beyond the pairwise level. 
For hypergraphs, the node degree $\ell$ (number of hyperedges connected to a node) and the number
of neighbors $k$ of a node differ from each other in contrast to the case of graphs, 
where counting the number of edges is equivalent to counting the number of neighbors.
In this article, I calculate the distribution of the number of node neighbors in random hypergraphs
in which hyperedges of uniform rank $r$ have a homogeneous (equal for all hyperedges) probability $p$ to appear.
This distribution is equivalent to the degree distribution of ensembles of graphs created as projections of 
hypergraph or bipartite network ensembles, where the projection connects any two nodes in
the projected graph when they are also connected in the hypergraph or bipartite network.
The calculation is non-trivial due to the possibility that neighbor nodes belong simultaneously to multiple
hyperedges (node overlaps). 
From the exact results, the traditional asymptotic approximation to the 
distribution in the sparse regime (small $p$) where overlaps are ignored is rederived and improved; the
approximation exhibits Poisson-like 
behavior accompanied by strong fluctuations modulated by power-law decays in the system size $N$ with decay exponents 
equal to the minimum number of overlapping nodes possible for a given number of neighbors.
It is shown that the dense limit cannot be explained if overlaps are ignored, and the correct
asymptotic distribution is provided.
The neighbor distribution requires the calculation of a new combinatorial coefficient $Q_{r-1}(k,\ell)$,
which counts the number of distinct labelled hypergraphs of $k$ nodes, 
$\ell$ hyperedges of rank $r-1$, and where every node is connected to at least one hyperedge.
Some identities of $Q_{r-1}(k,\ell)$ are derived and applied to the verification of normalization and the
calculation of moments of the neighbor distribution. 
\end{abstract}
\maketitle
\section{Introduction}
Fuelled by the recent
availability of digitized data from many sources, including social, technological, and natural 
systems, the scientific community has placed renewed interest into quantitative analysis of large datasets.
In this context, {\it complex networks theory} has emerged as one of the most active research areas providing
new analytical techniques~\cite{rev-Albert}. In essence, complex networks focuses on developing
understanding of a system from its representation as a collection of objects called nodes
and the relations between them, called edges. The set of nodes and edges together are known as a
graph (in mathematics) or network (in complex networks theory and in physics). 
Some examples of network representations are people and their friendships, particles and their collisions,
or statistical variables and their correlations.

The techniques of complex networks are meant to be quite general.
Some well studied examples of graphs are social networks~\cite{Onnela}, 
power grids~\cite{Crucitti}, and networks of infectious disease propagation~\cite{Colizza}, 
although there are many more systems that are being tackled with these techniques.
The general approach of complex networks is to study the statistical properties of a 
graph $\sigmabold$ or set of graphs $\sigmaconf$ such as degree distribution (where degree is the number 
of edges connected to a node, equivalent to the number of node neighbors), 
distribution of shortest path lengths among nodes (which is
at the core of the small world notion and of six degrees of separation~\cite{Milgram,Watts-Strogatz}), 
and community structure (loosely defined as groups of nodes among which 
there are more edges than with the rest of the graph)~\cite{Fortunato,Porter}. 
Of all these properties, the degree distribution is perhaps the most widely used in ongoing research,
due to its relevance in several other quantities such as the percolation threshold of a network~\cite{rev-Albert}.

In some systems, interactions occur in groups of nodes that may be larger than two. There are numerous
examples of this, such as the social networks in which infectious disease propagate, or the
statistical interactions between correlated events in financial systems.
Regardless of the context, when such multiway interactions occur it is convenient to use hypergraphs,
which generalize graphs by substituting edges with hyperedges, conglomerates of nodes that interact together
in groups of size $r$ (so-called hyperedge rank) $\geq 2$ (a simple graph or network is a specialization of
a hypergraph with $r=2$ exclusively).
Hypergraphs carry equivalent notions to those of graphs, such as path length and degree~\cite{Berger}.
This approach is gradually gaining attention~\cite{Ghoshal,Bradde,Newman-clusters,Wang}.

The degree of a node changes meaning slightly in hypergraphs. 
While degree continues to be the number of hyperedges a node is connected to, this is no
longer equivalent to the number of node neighbors a given node has. 
In the context of ensembles of random hypergraphs or graphs, as is our interest here, 
this change indicates that one must separately measure the node degree distribution and 
the node neighbor distribution.
This later quantity (henceforth referred to as neighbor distribution for short),
has received little direct attention despite its intuitive relevance (see illustrative discussion on disease propagation
at the end of Sec.~\ref{degree-dist-sec}, where the impact on quarantine numbers is discussed). 
In this article, I focus on the neighbor distribution in the case of homogeneous random hypergraphs 
of uniform rank $r$ (all hyperedges are of size $r$),
and derive complete results that cover all hypergraph densities. This is done via hypergraph
projections onto graphs as explained next~\cite{fn-complement}.

To determine the neighbor distribution in hypergraph ensembles, it is equivalent to look at projections of hypergraphs 
onto graphs and calculate the usual degree distribution in the projected graph ensembles~\cite{Lopez-projections}.
The projections are defined so that if two nodes are connected by any hyperedge then the projected graph 
has an edge between those nodes.
The notion of projection, useful here as a tool to calculate neighbor distribution,
is important in its own right because it is customary to first attempt to use graphs whenever possible,
typically weighted graphs, before introducing hypergraphs~\cite{Wasserman,Lopez-projections}.
It is worthwhile to point out that an equivalence can be established between hypergraphs 
and bipartite networks~\cite{Newman-WS} as explained in Chap. 7 of Ref.~\cite{Wasserman},
making this work useful in that context as well. 
For bipartite networks, the graph projection
corresponds to so-called one-mode networks, where once again, the degree distribution is the 
quantity of interest. Some relevant work has been done for bipartite networks that is related to the topic
of this article~\cite{Newman-WS,Ramasco,Nacher}, but it is confined to the sparse limit, and therefore
still leaves unanswered questions.

The complication in calculating the neighbor distribution is that it
is affected by a kind of degeneracy due to the potential presence of one or more nodes 
in multiple hyperedges (node overlaps). This makes the distribution calculation non-trivial.
In tracking this degeneracy, the need for a new enumerative quantity emerges.
If $k$ represents number of neighbors and $\ell$ number of hyperedges, the enumerative quantity is $Q_{r-1}(k,\ell)$
which, as explained below, is the cardinality of the set of 
all possible ways that $\ell$ hyperedges of rank $r$ anchored to a specific node visit
exactly $k$ distinct other nodes. $Q_{r-1}(k,\ell)$ also corresponds to the number of distinct 
labelled hypergraphs 
with $k$ nodes and $\ell$ hyperedges of rank $r-1$ such that all nodes belong to at least one hyperedge.
As far as the author is aware, this is the first study of $Q_{r-1}(k,\ell)$; some partial
results exist for the case of $r-1=2$ in Refs.~\cite{Bender-Canfield-McKay,Korshunov,OEIS}. 
In this article, $Q_{r-1}(k,\ell)$ is calculated by two different methods, and a number of identities 
relevant to the neighbor distribution are derived for it.
The calculation of $Q_{r-1}(k,\ell)$ allows for an exact solution to the neighbor distribution, as well as
the derivation of its sparse and dense asymptotics. 
In the conclusions, I briefly describe how to tackle the full problem where rank $r$ is no longer uniform.

A number of excellent recent publications~\cite{Ghoshal,Newman-clusters,Wang} touch on 
a related form of the neighbor distribution problem posed here, by counting neighbors multiple times if
they are part of different hyperedges. However, in those publications, the focus resides in the sparse limit,
where overlaps are small (see results in Sec.~\ref{degree-dist-sec}), and therefore the error made is asymptotically
small, decaying in inverse proportion to the system size.

The structure of the paper is as follows: Sec.~\ref{degree-dist-sec} focuses on constructing
the basics of hypergraph projections onto graphs, and showing the expressions for the neighbor
distribution of the projected graphs in general and in the dense and sparse limits. 
Section~\ref{Qr-sec} deals with the calculation of 
$Q_{r-1}(k,\ell)$ by two methods: inclusion-exclusion principle of combinatorics, and graph assembly. 
The later method is developed in detail for $r=3$ and additional results are developed to apply it to
$Q_{r-1}(k,\ell)$, i.e., general $r$. In order to apply $Q_{r-1}(k,\ell)$ to the neighbor
distribution, a number of combinatorial identities are derived and presented in Sec.~\ref{identities-sec}.
The conclusions are presented in Sec.~\ref{conclusions}.
\section{Hypergraph to graph projections and the calculation of the neighbor distribution}
\label{degree-dist-sec}
Consider a hypergraph $\sigmabold$ consisting of a set of nodes $1,\dots,N$,
and a set of hyperedges of rank $r$.
Each hyperedge has $r$ nodes $i_1,\dots,i_r$, and is assigned an indicator 
$\sigma_{i_1,\dots,i_r}$ equal to 1 if it is present in $\sigmabold$, and 0 if it is absent.
For simplicity, I focus on undirected hypergraphs (indicators
$\sigma_{i_1,\dots,i_r}$ are symmetric under permutations of $i_1,\dots,i_r$).
The hypergraphs are also homogeneous and non-interacting, where all hyperedges have equal 
probability $p$ to occur. 
Using the homogeneity and absence of interaction, the probability $P(\sigmabold)$ to observe configuration 
$\sigmabold$ is given by
\begin{equation}
P(\sigmabold)=p^{L(\sigmabold)}(1-p)^{{N\choose r}-L(\sigmabold)}
\label{Psigma_L}
\end{equation}
where $L(\sigmabold)$ is the number of hyperedges in $\sigmabold$. 
By defining $\TboldNr$ as the set of all possible hyperedges $\{(1,\dots,r),\dots,(N-r+1,\dots,N)\}$, 
the result above can also be written as
\begin{equation}
P(\sigmabold)=\prod_{(i_1,\dots,i_r)\in \TboldNr}p^{\sigma_{i_1,\dots,i_r}}(1-p)^{1-\sigma_{i_1,\dots,i_r}}
\label{Psigma_T}
\end{equation} 
where $\sigma_{i_1,\dots,i_r}$ are the hyperedges of $\sigmabold$.

\begin{figure}
\epsfig{file=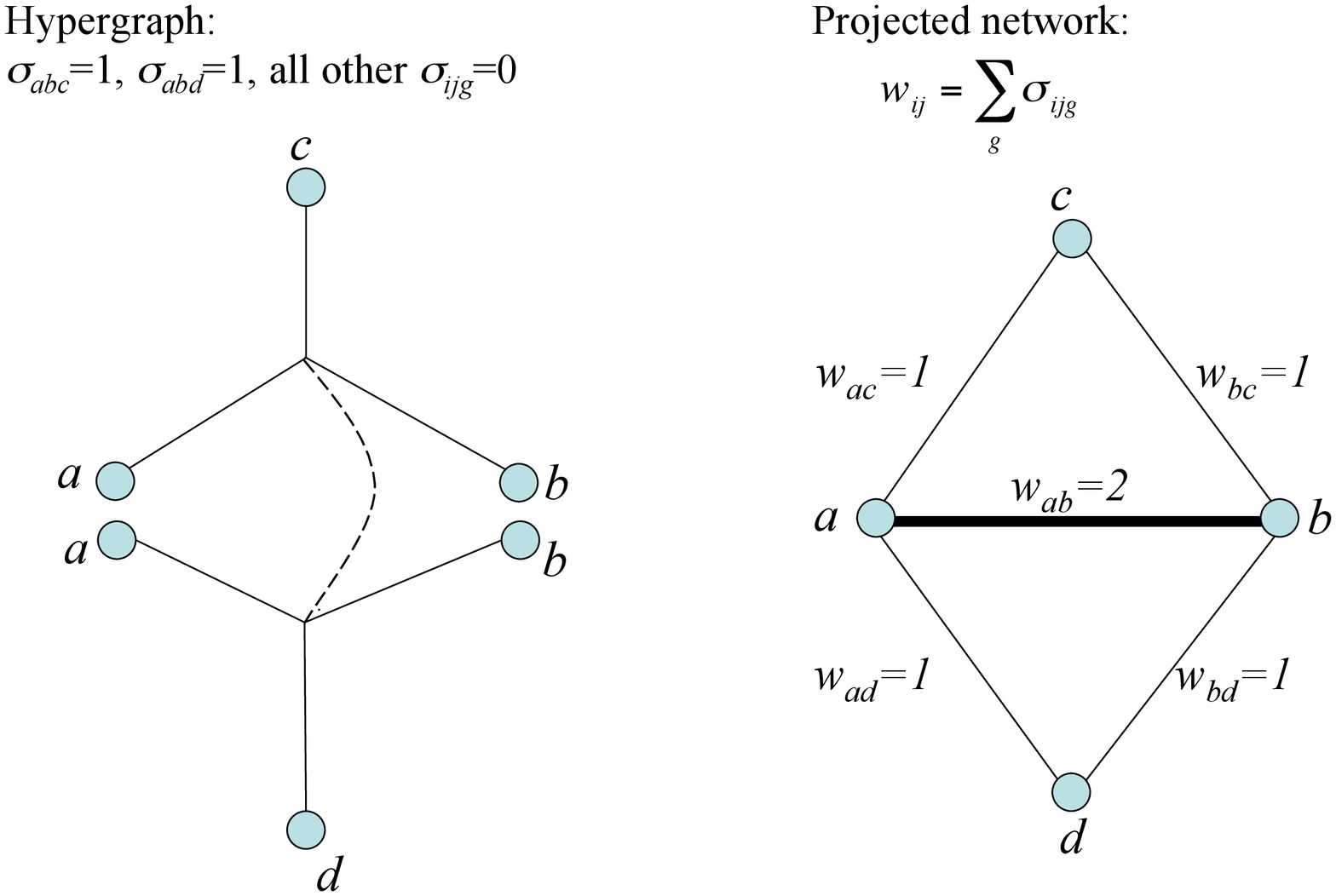,angle=270,scale=0.5}
\caption{Illustration for the projection $\mathcal{P}(o_{ij})=o_{ij}$ from hypergraphs $r=3$ to networks. 
On the left, the hypergraph is composed of hyperedges $\sigma_{a,c,b}=1$ and $\sigma_{a,b,d}=1$.
The projected network (right) has a link between $i$ and $j$ of weight $w_{ij}$ if there are $w_{ij}$ 
hyperedges that contain both nodes $i$ and $j$. In this example, $a$ and $b$ belong to two hyperedges, and 
thus $w_{a,b}=2$; all other node pairs belong in a single hyperedge, and thus their weights are equal to 1.
Note that $k_a=k_b=3$ instead of 4 because 3 is the number of {\it unique} neighbors each
of those nodes is connected to.}
\label{wij-project-illustration}
\end{figure}

The general hypergraph projection onto a graph~\cite{Lopez-projections} is defined as a function $\mathcal{P}$ applied
over the hyperedges of $\sigmabold$ that produces the adjacency matrix $w_{ij}$
for the projected weighted graph $G(\sigmabold)$. Each $w_{ij}$ is the indicator for edge $ij$ in $G$, 
but $w_{ij}$ can be any real positive number including zero,
making $G$ a weighted graph. $G(\sigmabold)$ is formed by the same node set as $\sigmabold$, 
together with edges that satisfy $w_{ij}>0$. 
Note that if a node does not belong to any hyperedge, it is isolated in both $\sigmabold$ and $G$.
For given $\sigmabold$, one can define the subset
$O_{ij}(\sigmabold):= \{(i_1,\dots,i_r)|(i_1,\dots,i_r)\in\sigmabold
\wedge i\in \{i_1,\dots,i_r\} \wedge j\in \{i_1,\dots,i_r\}\}$ of its hyperedges
that include simultaneously nodes $i$ and $j$.
It is natural to study projections of the type
\begin{equation}
w_{ij}(G)=\mathcal{P}(o_{ij}(\sigmabold)),
\label{wij-P}
\end{equation}
where $o_{ij}\equiv\left|O_{ij}(\sigmabold)\right|$ is the size (cardinality) of $O_{ij}(\sigmabold)$.
Thus, the weight of link $ij$ in $G$ only depends on the number of hyperedges that contain $i$ and $j$
(an intuitive choice, although certainly not the only possible model).
Furthermore, it is sensible to introduce the additional assumption that $w_{ij}>0$ iff $o_{ij}>0$,
or in other words, any pair of nodes $ij$ in the graph has non-zero weight if its corresponding $O_{ij}$ 
is not empty. An illustration of the projection process for the case $w_{ij}=\mathcal{P}(o_{ij})=o_{ij}$
and $r=3$ is shown in Fig.~\ref{wij-project-illustration}.

For projections as those defined above, the number of neighbors of node $i$ in $G(\sigmabold)$ is given by
\begin{equation}
k_i(\sigmabold)=\sum_{j=1;j\neq i}^N\theta(o_{ij}(\sigmabold))=\sum_{j=1;j\neq i}^N
\theta\left(\sum_{(i_1,\dots,i_r)\in O_{ij}(\sigmabold)}\sigma_{i_1,\dots,i_r}\right)
\label{ki_def}
\end{equation}
where $\theta(x)$ is the Heaviside step function, equal to zero if $x\leq 0$, and 1 if $x>0$.
To determine the neighbor distribution $\psi_i(k_i,p)$, one uses
\begin{multline}
\psi_i(k_i,p)=\sum_{\sigmabold\in\sigmaconf}
\delta\left(\sum_{j=1;j\neq i}^N\theta(o_{ij}(\sigmabold)),k_i\right) P(\sigmabold)\\
=\sum_{\sigma_{1,\dots,r}=0}^1p^{\sigma_{1,\dots,r}}(1-p)^{1-\sigma_{1,\dots,r}}\dots
\sum_{\sigma_{N-r+1,\dots,N}=0}^1 p^{\sigma_{N-r+1,\dots,N}}(1-p)^{1-\sigma_{N-r+1,\dots,N}}\\ 
\delta\left(\sum_{j=1;j\neq i}^N\theta(o_{ij}(\sigmabold)),k_i\right).
\label{psi_ki_def}
\end{multline}
where $\sigmaconf$ represents the set of all configurations contained in the 
homogeneous non-interacting hypergraph ensemble, and $\delta$ corresponds to a Kronecker delta.
Equation~(\ref{Psigma_T}) allows factorizing the sum over configurations in Eq.~(\ref{psi_ki_def}) to
produce the second line of the equation.
Only configurations $\sigmabold$ for which
$\sum_{j=1;j\neq i}^N\theta(o_{ij}(\sigmabold))$ is equal to $k_i$ contribute to $\psi_i(k_i,p)$. 
Only hyperedges where one of the indices $i_1,\dots,i_r$ is equal to $i$ are relevant to $k_i$; all
other hyperedges contribute the factor 
$\sum_{\sigma_{i_1,\dots,i_r}=0}^1p^{\sigma_{i_1,\dots,i_r}}(1-p)^{1-\sigma_{i_1,\dots,i_r}}=1$.
Let us label $\TboldNri$ the set of hyperedges that contribute to $k_i$ over all possible configurations.
As explained in the following, completing the calculation of $\psi_i(k_i,p)$ requires determining the terms in 
Eq.~(\ref{psi_ki_def}) that lead the delta to be 1, which is equivalent to finding all sets of hyperedges 
$(i_1,\dots,i_r)\in\TboldNri$ where $\sigma_{i_1,\dots,i_r}=1$, and the nodes involved in the set visit
exactly $k_i$ nodes as well as $i$.

\begin{figure}
\epsfig{file=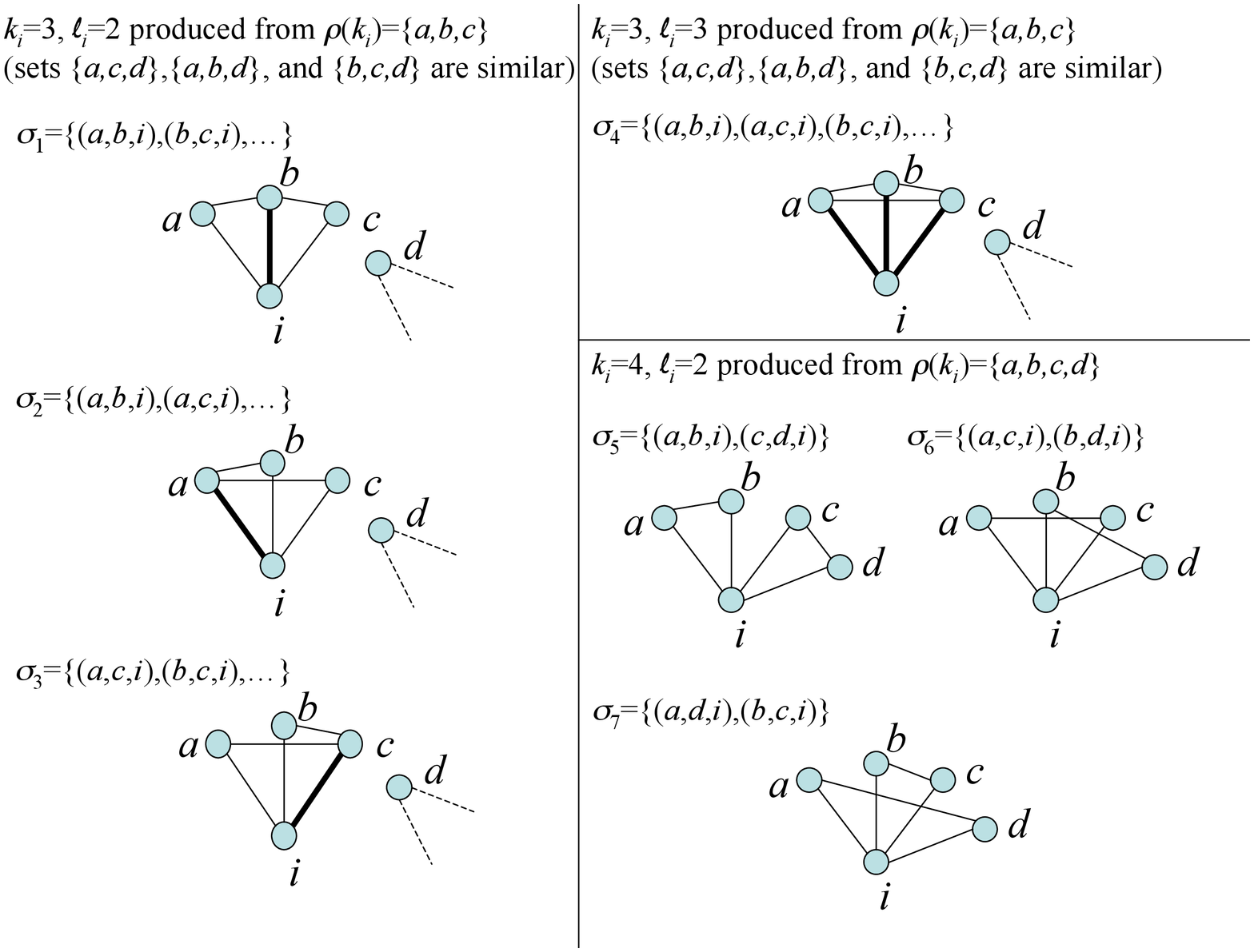,angle=270,scale=0.5}
\caption{Illustration ($r=3$) of the emergence of degree $k_i$ as a consequence of various possible 
hyperedge configurations. The figure also illustrates $Q_{r-1}(k_i,\ell_i)$. 
For $\rho(k_i)=\{a,b,c\}$ (left and top right panels), node $i$ can be connected to nodes 
$a,b,c$ in several ways. Hypergraphs
$\sigmabold_1,\sigmabold_2,\sigmabold_3$ exhibit the three ways in which $\ell_i=2$ hyperedges
can produce the connection between $i$ and all nodes of $\rho(k_i)$; $\sigmabold_4$ represents the 
single possibility of $\ell_i=3$ to connected $i$ to all nodes in $\rho(k_i)$. The successive dots
represent other hyperedges not connected to $i$, and hence irrelevant to $i$. From this example, 
$Q_{r-1=2}(k_i=3,\ell_i=2)=3$ and $Q_{r-1=2}(k_i=3,\ell_i=3)=1$. Note that in all 
hypergraphs $\sigmabold_1,\sigmabold_2,\sigmabold_3$ one node overlaps in two hyperedges and
in $\sigmabold_4$ all three nodes are overlapping in the appropriate pair of hyperedges.
For $\rho(k_i)=\{a,b,c,d\}$ the degree is $k_i=4$, and if $\ell_i=2$ as in the bottom right panel, 
one deduces that $Q_{r-1=2}(k_i=4,\ell_i=2)=3$ (the three hypergraphs 
$\sigmabold_5,\sigmabold_6,\sigmabold_7$).
}
\label{ki-illustration}
\end{figure}

The presence of the $\theta$ function in the definition of $k_i$ is the source of 
complications in the calculation of node neighbors. Thus, several configurations $\sigmabold$ can lead to
the same number of neighbors in a projected network. Figure~\ref{ki-illustration} illustrates the different possible
situations. From the figure, note the ways in which $k_i=3$ can emerge from various hypergraphs. 
All the possible hyperedge configurations that lead
to $k_i=3$ involving nodes $\{a,b,c,i\}$ (Fig.~\ref{ki-illustration} left and top right panels) are as follows: 
i) $(a,b,i)$ and $(b,c,i)$, ii) $(a,b,i)$ and $(a,c,i)$,
iii) $(a,c,i)$ and $(b,c,i)$, or iv) $(a,b,i)$, $(a,c,i)$, and $(b,c,i)$. The first three
possibilities have two hyperedges (denoted by $\ell_i=2$), and the last possibility has three hyperedges
($\ell_i=3$). 
If the hyperedges would involve another set of nodes, say $\{a,b,d,i\}$, a similar situation would occur.
As can be seen, in this example $k_i=3$ always occurs with some node neighbor appearing
in more than one hyperedge. This effect, referred here as {\it node overlaps}, is what
makes the calculation of $\psi_i(k_i,p)$ non-trivial; the $\theta$ function in Eq.~(\ref{psi_ki_def})
``deals'' with the overlaps.
Note that $\ell_i=2$ can also generate $k_i=4$ (e.g., bottom right of 
Fig.~\ref{ki-illustration}, where two hyperedges involve the nodes $\{a,b,c,d,i\}$, and there are no
overlaps).
All the cases just described play a role in the calculation of $\psi_i(k_i,p)$.

The examples above provide a way to proceed with the calculation. First, one can concentrate
on a specific set of $k_i$ nodes that connect to $i$, say $\rho(k_i)$, which guarantees that 
the degree is $k_i$ (the choice of $\rho(k_i)$ must be feasible, i.e., $k_i$ cannot be equal to 
$1,\dots,r-2$ for $r$-uniform hypergraphs).
Consider $\rho(k_i)=\{a,b,c\}$ and define $Q_{r-1}(k_i,\ell_i)$, the number of ways to achieve 
degree $k_i$ from set $\rho(k_i)$ using $\ell_i$ hyperedges of rank $r$. 
Hence, for $r=3$, $Q_2(k_i=3,\ell_i=2)=3$ (Fig.~\ref{ki-illustration} left panel) and $Q_2(k_i=3,\ell_i=3)=1$
(Fig.~\ref{ki-illustration} top right panel).
A second example is presented in Fig.~\ref{ki-illustration} (bottom right panel) for $\rho(k_i)=\{a,b,c,d\}$, 
producing $Q_2(k_i=4,\ell_i=2)=3$. The sub-index $r-1$ in $Q$ comes from the fact that each hyperedge
connected to $i$ also connects to $r-1$ other nodes which form an $r-1$-hyperedge with each other; for $r=3$,
as in Fig~\ref{ki-illustration}, these $r-1$-hyperedge are simply edges between nodes, such as
$(a,b)$, $(a,c)$, or $(b,c)$.

\begin{figure}
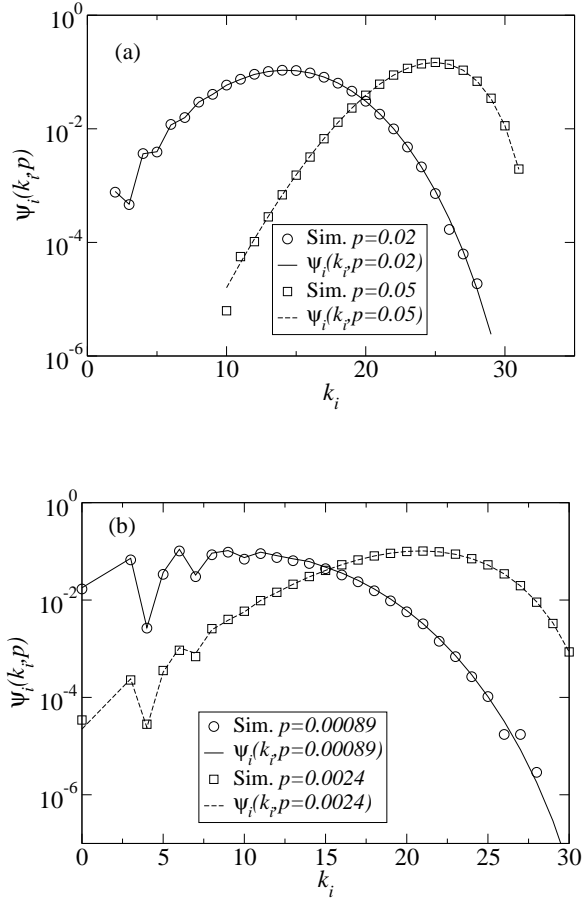

\epsfig{file=fig3a-v2.eps,scale=0.3}\\
\vspace{1cm}
\epsfig{file=fig3b-v2.eps,scale=0.3}
\caption{$\psi_i(k_i,p)$ vs. $k_i$ calculated both from theory Eq.~(\ref{psi-ki-final}) (curves) and simulations
(symbols):
(a) $N=32$, $r=3$, and $p\approx 0.02~(\bigcirc),0.05~(\square)$ with $10^5$ realizations per simulation.
(b) $N=32$, $r=4$, and $p\approx 0.00089~(\bigcirc),0.0024~(\square)$ with $10^4$ realizations per simulation.
}
\label{psik-figure}
\end{figure}

\begin{figure}
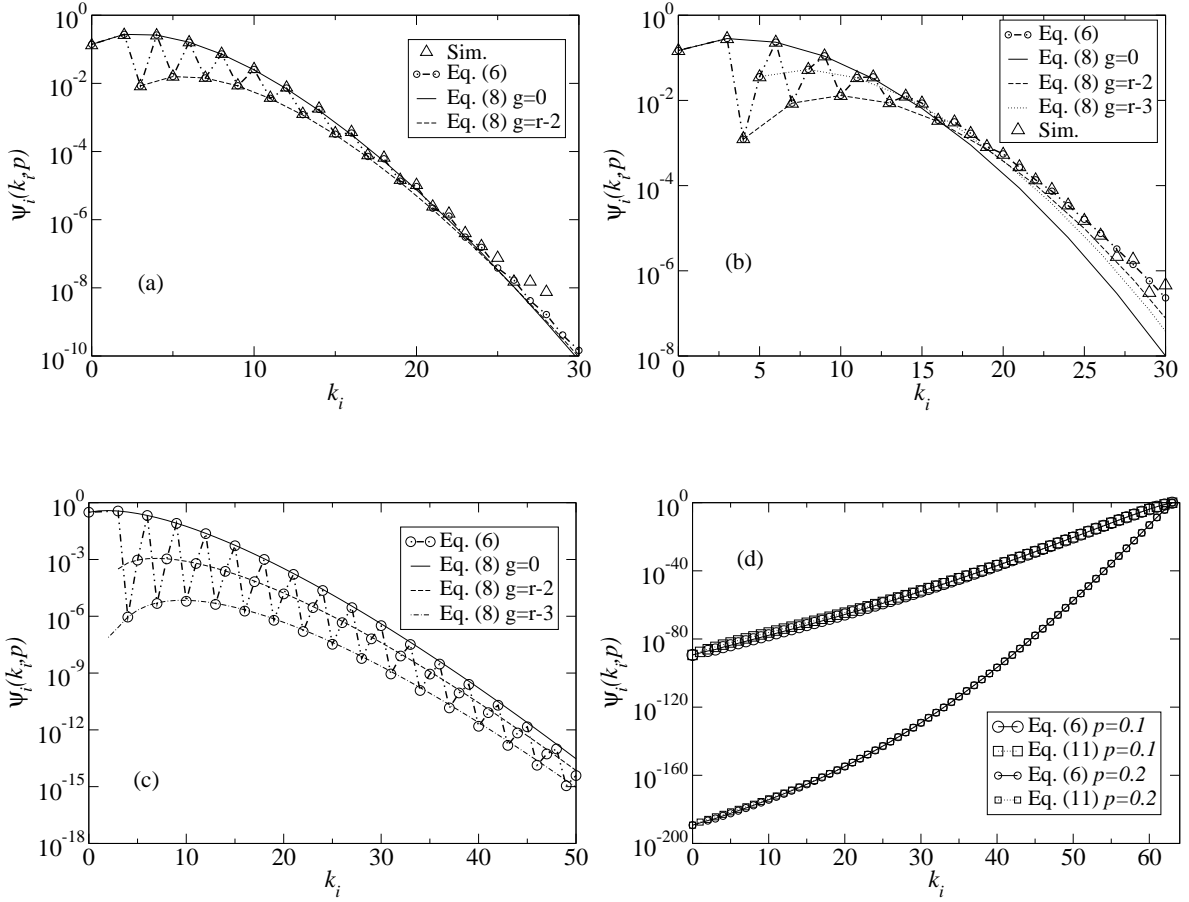

\epsfig{file=fig4a-v2.eps,scale=0.3}
\vspace{1cm}
\epsfig{file=fig4b-v2.eps,scale=0.3}
\epsfig{file=fig4c-v2.eps,scale=0.3}
\epsfig{file=fig4d-v2.eps,scale=0.3}
\caption{$\psi_i(k_i,p)$ vs. $k_i$ from Eq.~(\ref{psi-ki-final}) in the sparse limit (close to percolation),
and the sparse approximations $\psi_i(k_i,p,g)$ given in Eq.~(\ref{psi-poisson}):
(a) $N=128, r=3$ and $p$ is adjusted to $\langle k\rangle=4$ (simulations have $10^6$ realizations), 
(b) $N=64, r=4$ and $p$ is adjusted to $\langle k\rangle=6$ ($10^5$ realizations), and
(c) $N=2048, r=4$ and $p$ is adjusted to $\langle k\rangle=3.5$ (all curves are theoretical).
In all plots, the values from Eq.~(\ref{psi-ki-final}) are represented by ($\bigcirc$) connected with 
the double dot-dashed line,
simulations by ($\triangle$), and the approximations $\psi_i(k_i,p,g)$ by thin continuous line for $g=0$,
dashed line for $g=r-2$, and dot-dashed line for $g=r-3$. For small systems such as (a) and (b), it
is better to use Eq.~(\ref{psi-sparse}) for $\psi_i(k_i,p,g)$, since the asymptotic limit is still not fully
expressed.
However, in (c) the size is large enough and displays good asymptotics, given by Eq.~(\ref{psi-poisson}).
(d) $\psi_i(k_i,p)$ vs. $k_i$ in the dense limit for $N=64,r=3$ and $p=0.1,0.2$ from Eq.~(\ref{psi-ki-final})
($\bigcirc$) and the dense approximation Eq.~(\ref{psi-dense})($\square$) (all theoretical results). 
As $p$ increases the agreement improves.
}
\label{psi-sparse-figure}
\end{figure}

Applying the ideas of the previous paragraph, one can determine that the contribution to 
$\psi_i(k_i,p)$ from a specific set $\rho(k_i)$ of
nodes {\it and} number of hyperedges $\ell_i$ is given by $Q_{r-1}(k_i,\ell_i)p^{\ell_i} 
(1-p)^{{N-1\choose r-1}-\ell_i}$, where ${N-1\choose r-1}$ comes from
the size of $\TboldNri$. Note also that $\ell_i$ must satisfy some constraints for given $k_i$:
in order to be able to visit $k_i$ nodes, the smallest number of hyperedges necessary
is $\lceil k_i/(r-1)\rceil\leq\ell_i$, where $\lceil .\rceil$ is the ceiling function; also, 
there are ${k_i\choose r-1}$
ways to choose node groups of size $r-1$ out of $k_i$ nodes, and thus $\ell_i\leq{k_i\choose r-1}$. 
Therefore, conditional on $\rho(k_i)$, 
$\psi_i(k_i,p|\rho(k_i))=\sum_{\ell_i=\lceil k_i/(r-1)\rceil}^{k_i\choose r-1} Q_{r-1}(k_i,\ell_i)
p^{\ell_i} (1-p)^{{N-1\choose r-1}-\ell_i}$. The final step is to note that there
are ${N-1\choose k_i}$ ways to select $\rho(k_i)$, leading to~\cite{Lopez-projections} 
\begin{equation}
\psi_i(k_i,p)={N-1\choose k_i}\sum_{\ell_i=\lceil k_i/(r-1)\rceil}^{k_i\choose r-1}
Q_{r-1}(k_i,\ell_i)p^{\ell_i}(1-p)^{{N-1\choose r-1}-\ell_i}.
\label{psi-ki-final}
\end{equation}
Figure~\ref{psik-figure} shows examples of $\psi_i(k_i,p)$ from analytics and simulations, 
for the general case (``intermediate'' $p$), and $r=3,4$; Fig.~\ref{psi-sparse-figure} does the same
for the sparse and dense cases (small and large $p$).

In the sparse case, close to the percolation threshold of the hypergraphs,
large fluctuations appear in the distribution at relatively small $k_i$.
This behavior emerges because, at small $p$, the likelihood that hyperedges share multiple nodes (node overlaps) 
is low, which occurs when $k_i$ is not a multiple of $r-1$. 
To explain this, consider the low density regime when 
$p\sim\alpha p_c$ with $\alpha$ a constant of order $\gtrsim 1$ ($\alpha=1$ is the percolation threshold
as derived in Ref.~\cite{Lopez-projections}, 
with $p_c=\left(N\frac{\partial^2}{\partial N^2}{N\choose r}\right)^{-1}$). 
In this regime, $\psi_i(k_i,p)$ can in fact be well approximated by using only $\ell=\lceil k_i/(r-1)\rceil$, i.e.
\begin{equation}
\psi_i(k_i,p)\approx {N-1\choose k_i}Q_{r-1}\left(k_i,\left\lceil\frac{k_i}{r-1}\right\rceil\right)
p^{\left\lceil\frac{k_i}{r-1}\right\rceil}
(1-p)^{{N-1\choose r-1}-\left\lceil\frac{k_i}{r-1}\right\rceil};
\qquad [p\sim\alpha p_c].
\label{psi-sparse}
\end{equation}
The direct calculation of $Q_{r-1}(k_i,\lceil k_i/(r-1)\rceil)$ is addressed in Sec.~\ref{Qrmin}.
To track whether $k_i$ is a multiple of $r-1$, one can introduce $g\equiv\text{mod}(k_i,r-1)$,
where $0\leq g\leq r-2$.
If $g=0$, $k_i$ is a multiple of $r-1$. On the other hand, when $g\neq 0$, there are $r-1-g$ node overlaps.
For very large $N$, $p_c\sim \alpha (r-2)!/N^{r-1}$, which together with
Stirling's approximation and $Q(k_i,\lceil k_i/(r-1)\rceil)$ from Sec.~\ref{Qrmin}, 
lead to $\psi_i(k_i,p)\to \psi_i(k_i,p,g)$ in the sparse limit
\begin{equation}
\psi_i(k_i,p,g)\approx N^{g-(r-1)}
\frac{e^{-\alpha/(r-1)}}{\left(\left\lceil\frac{k_i}{r-1}\right\rceil\right)!}
\left(\frac{\alpha}{r-1}\right)^{\lceil k_i/(r-1)\rceil}X_g\left(r,\left\lceil\frac{k_i}{r-1}\right\rceil\right)
\qquad \left[p\sim\frac{\alpha(r-2)!}{N^{r-1}}\right],
\label{psi-poisson}
\end{equation}
where
\begin{equation}
X_g\left(r,\left\lceil\frac{k_i}{r-1}\right\rceil\right)
=\left\{
\begin{array}{ll}
1;&\qquad g=0\\
\frac{(r-1)^2}{2}\left\lceil\frac{k_i}{r-1}\right\rceil \left(\left\lceil\frac{k_i}{r-1}\right\rceil-1\right);&\qquad g=r-2\\
\vdots&\qquad\vdots\\
X_1\left(r,\left\lceil\frac{k_i}{r-1}\right\rceil\right);&\qquad g=1.
\end{array}
\right.
\end{equation}
Also, for the purposes of these approximations, one takes
\begin{equation}
\left\lceil\frac{k_i}{r-1}\right\rceil
=\left\{
\begin{array}{ll}
\frac{k_i}{r-1};&\qquad g=0\\
\frac{k_i+1}{r-1};&\qquad g=r-2\\
\vdots&\qquad\vdots\\
\frac{k_i+r-1-g}{r-1};&\qquad g=1,
\end{array}
\right.
\end{equation}
which give the correct value of $\lceil k_i/(r-1)\rceil$ for the specific $g$ listed.
Equation~(\ref{psi-poisson}) is quite informative. When $g=0$, the degree distribution is strictly Poisson, but when
$g\neq 0$, an asymptotic attenuation factor of the form $N^{g-(r-1)}$ appears, which indicates that the 
probability to observe a single node overlap ($1=r-1-g=r-1-(r-2)$) is reduced by a $1/N$ factor, a
2 node overlap ($2=r-1-g=r-1-(r-3)$)
by $1/N^2$, etc. The qualitative relevance of this result is that approximations of hypergraphs that 
consider the hyperedges as non-overlapping when projected onto a graph 
(or made into a one-mode network of a bipartite network) incur an error of order $N^{-1}$ 
in $\psi_i(k_i,p)$ in the sparse limit.
In Fig.~\ref{psi-sparse-figure}(a), (b) and (c), the actual distribution (as given by Eq.~(\ref{psi-ki-final}))
is plotted against simulations, and the curves
of Eq.~(\ref{psi-poisson}) are superimposed for confirmation; one plot is performed on a system size
much larger than those available for Monte Carlo simulation, but shows the best adherence to asymptotics. 
The case $g=0$ is an envelope for the distributions when $N\to\infty$.

\begin{figure}
\epsfig{file=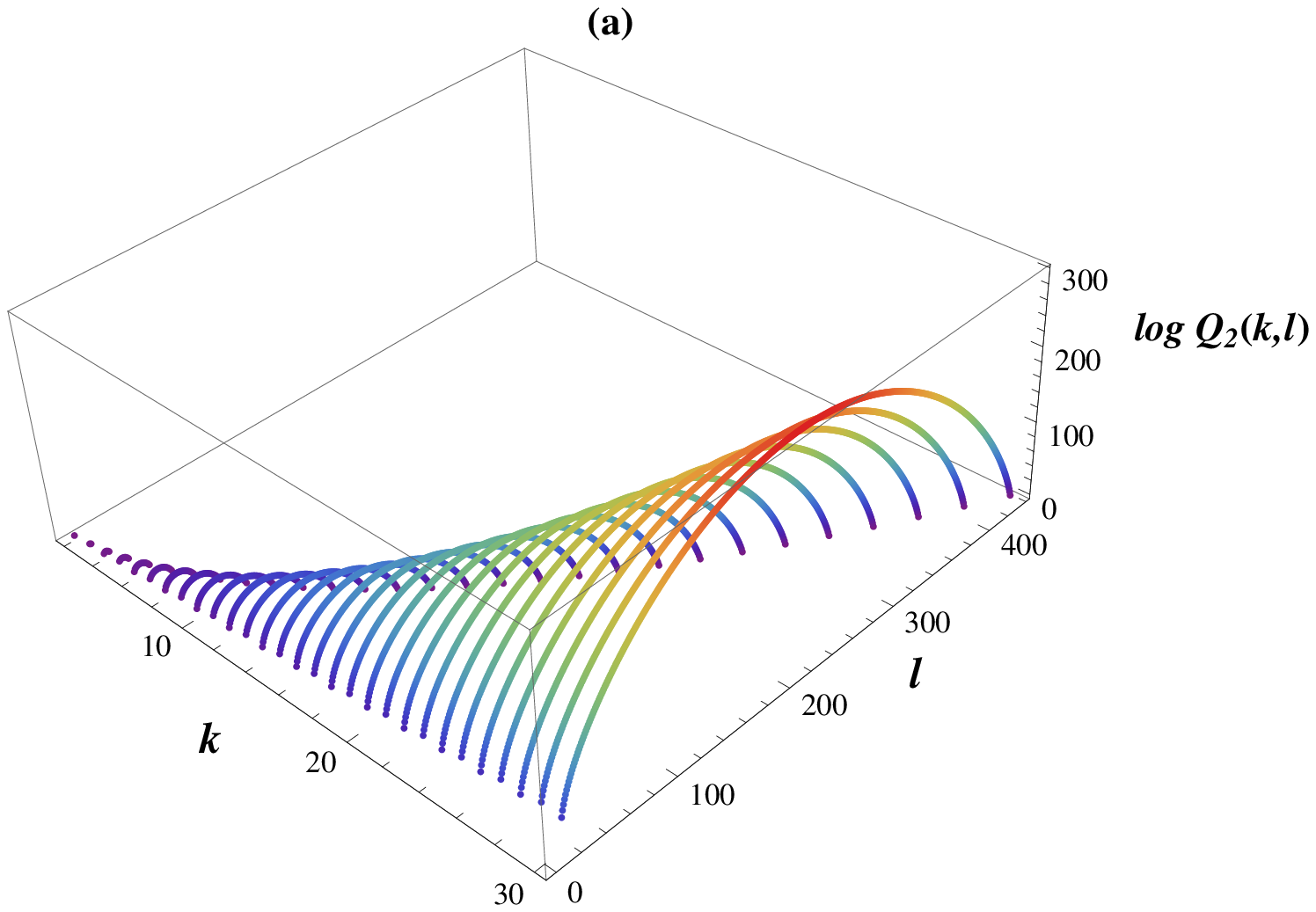,scale=0.7,bbllx=75,bblly=56,bburx=519,bbury=374,clip=}
\epsfig{file=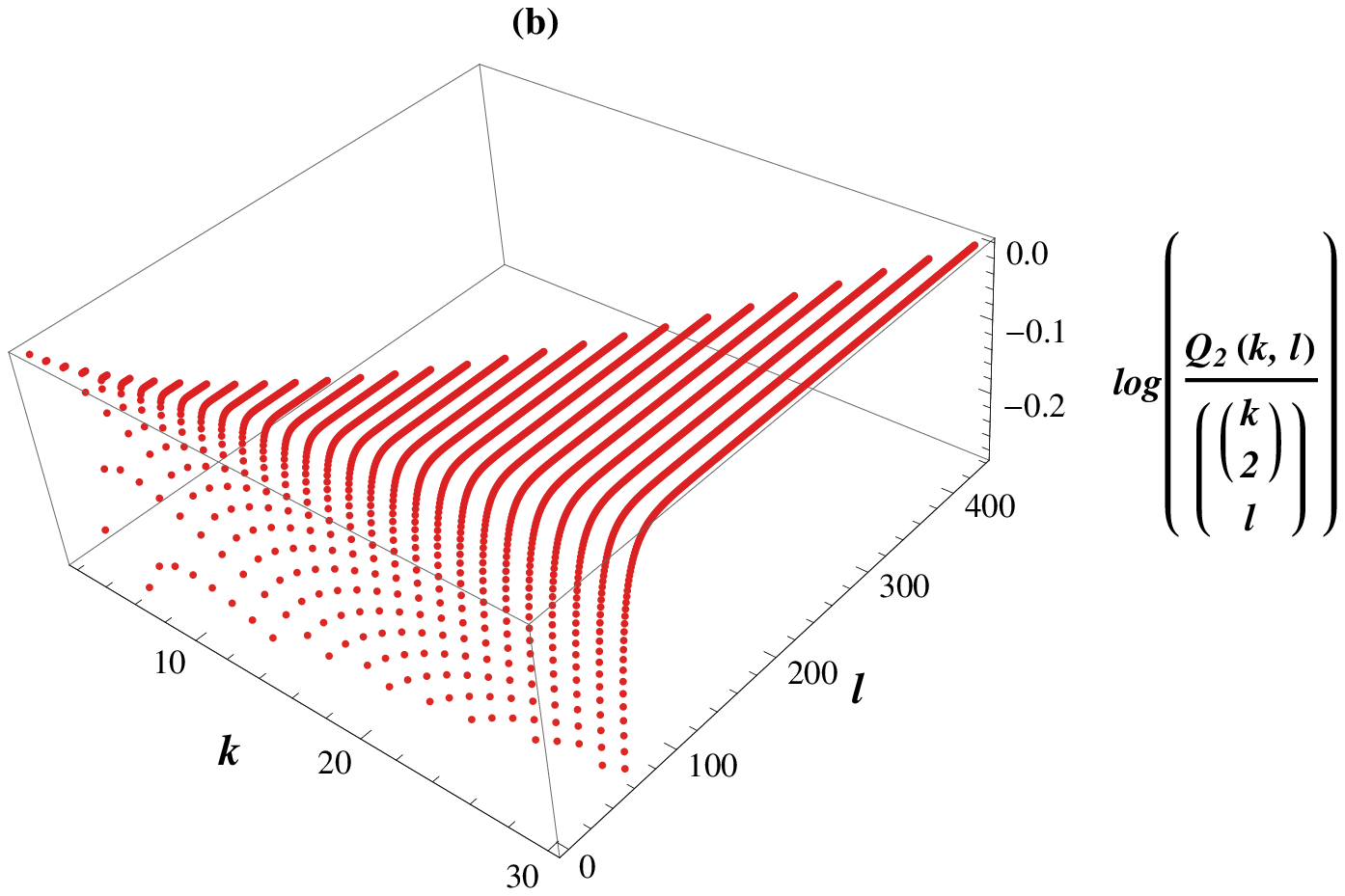,scale=0.8,bbllx=75,bblly=129,bburx=519,bbury=414,clip=}
\caption{(color online) (a) $\ln Q_2(k,\ell)$ vs. $k,\ell$ and (b) $\ln[Q_2(k,\ell)/{{k\choose 2}\choose\ell}]$ 
vs. $k,\ell$. From (b) it is clear that as $\ell$ increases for any $k$, the ratio tends to 1 (with $\log$ going to 0).
}
\label{Q2kl-asymptotic}
\end{figure}

In the dense limit, if overlaps are ignored when trying to estimate $\psi_i(k_i,p)$, the
error becomes overwhelming. To illustrate this, note that as $p\to 1$ the number of hyperedges
visiting $i$ approaches ${N-1\choose r-1}$, and with no overlaps this would lead to a
$k_i$ approaching $(r-1){N-1\choose r-1}$ which is clearly wrong as in reality $k_i$ can be at most $N-1$.
With the results in Sec.~\ref{Q2-calculation} and, in particular, the realization that for large $\ell_i$,
$Q_{r-1}(k_i,\ell_i)$ can be approximated as ${{k_i\choose r-1}\choose \ell_i}$ (see
Fig.~\ref{Q2kl-asymptotic}(b)),
the simple approximation 
\begin{equation}
\psi_i(k_i,p)\approx {N-1\choose k_i}(1-p)^{{N-1\choose r-1}-{k_i\choose r-1}}\qquad [\text{large }p]
\label{psi-dense}
\end{equation}
for finite and relatively large $p$ becomes satisfactory. This can be obtained by algebraic manipulation and the use
of the gaussian approximation for the summand of Eq.~(\ref{psi-ki-final}). Note that the limit $p\to 1$
is correctly obtained: for all $k_i<N-1$, the exponent of $1-p$ is positive, and as $p$
approaches 1, $\psi_i(k_i,p)\to 0$; only $k_i=N-1$ makes the exponent of $1-p$ equal to zero, 
producing the result $\psi_i(k_i=N-1,p=1)\to {N-1\choose N-1}=1$. 
Figure~\ref{psi-sparse-figure}(d) shows examples of the dense estimate
Eq.~(\ref{psi-dense}) against Eq.~(\ref{psi-ki-final}), which agree well with each other.

To fully specify Eq.~(\ref{psi-ki-final}), it is necessary to determine $Q_{r-1}(k_i,\ell_i)$.
In order to achieve this, it is important to develop some intuition about the meaning of
$Q_{r-1}(k_i,\ell_i)$. The case $r=3$ is very useful. Each hyperedge (in this case a triplet)
connects $i$ to two other nodes taken out of $\rho(k_i)$, and clearly
all nodes in $\rho(k_i)$ are visited at least once so that the degree is equal to $k_i$.
On any two nodes of $\rho(k_i)$, say $a$ and $b$, the 3-hyperedge that connects them to $i$ acts as an 
edge between $a$ and $b$.
Given that there are $\ell_i$ hyperedges available to achieve $k_i$, {\it determining 
$Q_2(k_i,\ell_i)$ is equivalent to enumerating all distinct labelled graphs of 
$k_i$ nodes and $\ell_i$ edges, in which all nodes have degree at least one; 
there are no isolated nodes}. Henceforth, I refer to these graphs as {conditioned graphs}.
In the examples in Fig.~\ref{ki-illustration}, 
the cases contributing to $k_i=3$ and $\ell_i=2$ are: i) $(a,b)$ and $(b,c)$, ii)
$(a,b)$ and $(a,c)$, and iii) $(a,c)$ and $(b,c)$, 
and to $k_i=3$ and $\ell_i=3$ is $(a,b)$, $(a,c)$ and $(b,c)$. 
When the problem is generalized, 
$Q_{r-1}(k_i,\ell_i)$ corresponds to the number of distinct labelled hypergraphs 
with $k_i$ nodes and $\ell_i$ hyperedges of rank $r-1$ such that all nodes 
belong to at least one hyperedge~\cite{fn-I}. 
In the next section, the calculation of $Q_2(k_i,\ell_i)$ is tackled through different techniques, 
leading to the two formulas (Eqns.~(\ref{Qr-inex}) and (\ref{Q2final}) where the first one is valid for all 
$r$).

To conclude this section and relate the model to some concrete applications, I determine $\langle k_i\rangle$
and explain its significance in a practical example, which also highlights the importance of
the exact results derived here.
One can calculate $\langle k_i\rangle$ using $P(\sigmabold)$ (later on, 
this calculation is repeated using $\psi_i(k_i,p)$ and identities relevant to $Q_{r-1}(k_i,\ell_i)$).
By definition
\begin{equation}
\langle k_i\rangle=\sum_{\sigmabold\in\sigmaconf}k_i(\sigmabold)P(\sigmabold)
=\sum_{\sigmabold\in\sigmaconf}\sum_{j=1;j\neq i}^N\theta(o_{ij}(\sigmabold))P(\sigmabold)
=\sum_{j=1;j\neq i}^N\sum_{\sigmabold\in\sigmaconf}\theta(o_{ij}(\sigmabold))P(\sigmabold).
\end{equation}
Concentrating on the sum over $\sigmabold$
\begin{equation}
\sum_{\sigmabold\in\sigmaconf}\theta(o_{ij}(\sigmabold))P(\sigmabold)
=\sum_{\sigmabold\in\sigmaconf}P(\sigmabold)-\sum_{o_{ij}(\sigmabold)=0}P(\sigmabold)
=1-\sum_{o_{ij}(\sigmabold)=0}P(\sigmabold),
\end{equation}
where one uses the realization that $\theta(o_{ij}(\sigmabold))=1$ in all hypergraphs where
$o_{ij}\geq 1$, and 0 if $o_{ij}=0$. To determine the last sum, one uses the independence of the
hyperedges in Eq.~(\ref{Psigma_T}), and therefore
\begin{multline}
\sum_{o_{ij}(\sigmabold)=0}P(\sigmabold)=
\sum_{\sigma_{1,\dots,r}=0}^1 p^{\sigma_{1,\dots,r}}(1-p)^{1-\sigma_{1,\dots,r}}\dots
\sum_{\sigma_{N-r+1,\dots,N}=0}^1 p^{\sigma_{N-r+1,\dots,N}}(1-p)^{1-\sigma_{N-r+1,\dots,N}}\\
=(1-p)^{N-2\choose r-2},
\end{multline}
where all hyperedges $\sigma_{i_1,\dots,i_r}=0$ when both $i$ and $j$ are among the indices so that
$o_{ij}=0$ (there are ${N-2\choose r-2}$ such hyperedges), and the sums over all
other $\sigma_{i_1,\dots,i_r}$ produce factors of 1. Since this result is independent of $j$, 
\begin{equation}
\langle k_i\rangle=\sum_{j=1;j\neq i}^N\left[1-(1-p)^{N-2\choose r-2}\right]
=(N-1)\left[1-(1-p)^{N-2\choose r-2}\right].
\label{avgki}
\end{equation}
Higher moments can also be calculated this way, but they introduce couplings among indices, and the previous
approach becomes much harder. 
In Sec.~\ref{identities}, a more powerful approach is developed making more straightforward the calculation
of higher moments.
Note that the low density regime $p=\alpha p_c$ corresponds to $\langle k_i\rangle\sim\alpha$.

\begin{figure}
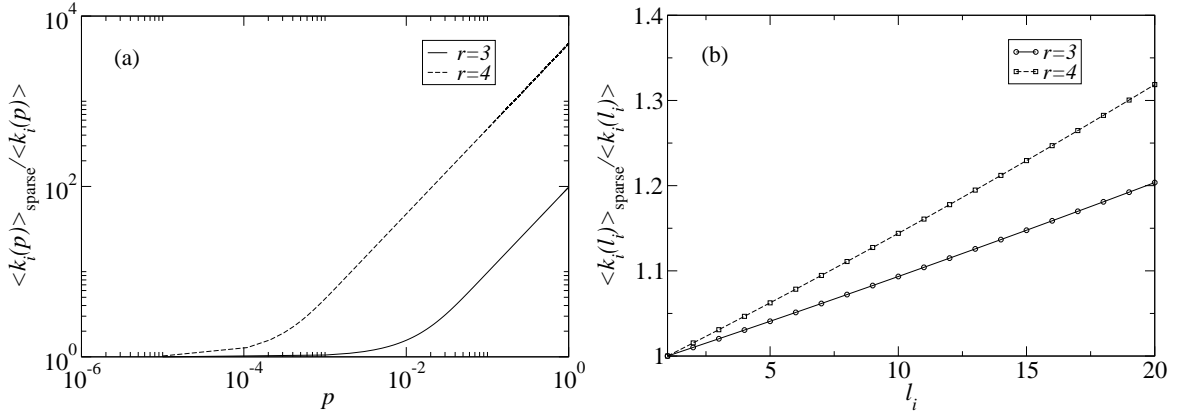

\epsfig{file=fig5a-v2.eps,scale=0.3}
\epsfig{file=fig5b-v2.eps,scale=0.3}
\caption{(a) Ratio $\langle k_i(p)\rangle_{\rm sparse}/\langle k_i(p)\rangle$ versus $p$ calculated
via the sparse approximation Eq.~(\ref{psi-poisson}) with $g=0$ and Eq.~(\ref{avgki}) for $N=100$. One can quickly
see the two quantities deviate considerably even for small $p$.
(b) Ratio $\langle k_i(\ell_i)\rangle_{\rm sparse}/\langle k_i(\ell_i)\rangle$ versus $\ell_i$
for $N=100$ and $r=3$ and 4,
which also shows how rapidly the sparse approximation and the exact results deviate from one another.}
\label{ksparse_k_ratio}
\end{figure}

To illustrate the relevance of the model in practice, consider the determination of quarantine 
levels necessary to isolate an individual with an infectious disease. To a first approximation,
the quantity of interest here is $\langle k_i\rangle$. In a traditional approach
with sparse approximate mathematical models, there would be considerable overestimations of quarantine levels
because node overlaps are ignored, and thus a friend or colleague that is part of two communities
at the same time would be counted twice. When the correct approach is taken, quarantine levels are estimated
in a more realistic way. In Fig.~\ref{ksparse_k_ratio}(a), I present the exact value of
the ratio $\langle k_i(p)\rangle_{\rm sparse}/\langle k_i(p)\rangle$ where $\langle k_i(p)\rangle$ is given by
Eq.~(\ref{avgki}) and $\langle k_i(p)\rangle_{\rm sparse}=(r-1){N-1\choose r-1}p$ which is the average number
of hyperedges connected to node $i$ times the $r-1$ neighbors that each edge contributes.
This ratio of averages, which is a measure of how much the sparse approximation differs from the correct value, 
deviates from 1 very rapidly even for a very small $p$. 
For a ratio $\langle k_i(p)\rangle_{\rm sparse}/\langle k_i(p)\rangle$ equal to 2, with $N=100$,
$p\approx 0.016$ for $r=3$ and $\approx 3.35\times 10^{-4}$ for $r=4$, which corresponds to
an expected number of hyperedges $\langle \ell\rangle\approx 79.3$ for $r=3$ and 52.60 for $r=4$.
On a social network this is a very small number of hyperedges, and thus in a quarantine situation, 
even at this sparse density, the sparse approximation fails suggesting twice as many individuals need to be
quarantined than when overlaps are considered.
Another way to measure the discrepancy in quarantine levels is to 
calculate $\langle k_i(\ell_i)\rangle_{\rm sparse}/\langle k_i(\ell_i)\rangle$,
which compares for a given node the number of neighbors $k_i$ expected in the sparse approximation and in
our calculations when the number of hyperedges $\ell_i$ connected to the node
is given; $\langle k_i(\ell_i)\rangle_{\rm sparse}=(r-1)\ell_i$ whereas
$\langle k_i(\ell_i)\rangle=(N-1)\left(1-{{N-2\choose r-1}\choose\ell_i}/{{N-1\choose r-1}\choose\ell_i}\right)$
which can be calculated from the ratio between Eqs.~(\ref{ki_Q_sum}) and (\ref{Q_sum_inverse}) in Sec.~\ref{identities}.
Here (see Fig.~\ref{ksparse_k_ratio}(b)), the ratio also moves away from 1 quickly, 
and by $\ell_i=20$ it overestimates the number of neighbors by $20\%$ for $r=3$ and $32\%$ for $r=4$.
These examples show that, essentially, in order to properly account for node neighbors in systems
with group structure, node overlaps can hardly be ignored, and the approach presented here becomes
necessary. 

Our random hypergraph model (considering also results in Ref.~\cite{Lopez-projections}) has
other domains of application, such as being a source of random null models for studies of data-constructed
networks. To take an example, if one considers a network structure given directly by
data, such as a metabolic network, certain structural features of the network can be compared to
random null models of the network to determine if they are statistically rare. 
If so, such features are potentially relevant biologically and may warrant further study. 
Furthermore, if the data-constructed model has a one-mode network representation
believed to be a useful simplification of the full hypergraph or bipertite network
model (e.g., because it lends itself to the use of some technique best
defined only on graphs), our framework provides the most complete
way to determine the statistics of the associated one-mode random null model,
and hence would prove useful for intepretation and analysis in this approach.
\section{Calculations of $Q_{r-1}(k,\ell)$}
\label{Qr-sec}
To determine $Q_{r-1}(k_i,\ell_i)$, I proceed by focusing on the enumeration of the conditioned graphs/hypergraphs 
mentioned above. 
{\it To avoid confusion, it is important to emphasize that the graphs and hypergraphs considered in this 
section are not those in $\sigmaconf$, but instead are tools to determine $Q_{r-1}$ and, if desired, can
be interepreted directly in the context of }$\sigmaconf$~\cite{fn-I}, {\it but it is not necessary.}
The nodes are labelled, consistent with the selection of sets $\rho(k_i)$ which are also composed of labelled
nodes. In the calculations of this section, given a choice $\rho(k_i)$ with $k_i$ nodes and $\ell_i$ 
hyperedges of rank $r-1$, the node $i$ is irrelevant and therefore the subindex $i$ is dropped.
\subsection{Inclusion-Exclusion formula}
The combinatorial coefficient $Q_{r-1}(k,\ell)$ can be determined
via the inclusion-exclusion principle of combinatorics~\cite{Riordan}. The idea behind this principle is
to count the number of elements in a set that satisfies certain conditions through a series of
alternative overcounts and undercounts.
Focusing on $Q_2(k,\ell)$
as the enumeration of conditioned graphs, a simple overcount of the conditioned graphs is
${{k\choose 2}\choose \ell}$, the number of graphs with $k$ nodes and $\ell$ edges, where there are ${k\choose 2}$
places to locate $\ell$ edges.
This overcounts $Q_2(k,\ell)$ because it ignores the condition of all nodes being connected to at least one edge. 
If the configurations in which at least one node is not connected are taken away from the previous enumeration, the
correct result is obtained. To approach this, one makes a first correction by taking away 
${k\choose k-1}{{k-1\choose 2}\choose\ell}$, which counts all choices of $k-1$ nodes picked out of $k$ multiplied by
the number graphs formed with $k-1$ nodes and $\ell$ edges. This step has now eliminated all configurations that
have nodes disconnected, but has eliminated multiple times all graphs in which two or more nodes are not connected
to an edge. To correct for this, it is necessary to add ${k\choose k-2}{{k-2\choose 2}\choose\ell}$.
Once again there are unwanted graphs in this count which require further correction. It is straightforward to
continue this until the point when the choice of $m$ nodes chosen out of $k$ is small enough that 
${m\choose 2}<\ell$, at which point the sequence stops. These considerations lead to the expression
\begin{equation}
Q_{2}(k,\ell)=\sum_{m=0}^{k}(-1)^{k-m}{k\choose m}{{m\choose 2}\choose\ell}.
\label{Q2-inex}
\end{equation}
The extension to arbitrary $r$ is direct, producing
\begin{equation}
Q_{r-1}(k,\ell)=\sum_{m=0}^{k}(-1)^{k-m}{k\choose m}{{m\choose r-1}\choose\ell}.
\label{Qr-inex}
\end{equation}
In terms of direct calculation, this formula is useful in producing an exact numerical result, but it is 
not so easy to interpret on the basis of $k$ and $\ell$, and some calculations that depend
on it become difficult due to the alternating signs (e.g. asymptotics).  
\subsection{Assembly of $Q_2(k,\ell)$}
\label{Q2-assembly}
An alternative to inclusion-exclusion is that of graph assembly.
In this section, I explain how to compute $Q_{r-1}(k,\ell)$ with $r=3$ through assembly. 
The extension to arbitrary $r$ is explained in Sec.~\ref{Qr-extension}, and though it is straightforward,
it is admittedly cumbersome.
Nevertheless, the picture developed here is more intuitive than inclusion-exclusion, and opens the possibility 
to study the properties of $Q_{r-1}(k,\ell)$ further.
To develop the procedure to count assemblies leading to the conditioned graphs, small examples
are presented where $\ell$ is close to its minimum possible value for given $k$. 
These examples exhibit 
all the aspects necessary to deal with the general $Q_2(k,\ell)$ case, which is studied
in Sec.~\ref{Q2-calculation}.
\subsubsection{Preliminaries and simple examples. Types of edges}
\begin{figure}
\epsfig{file=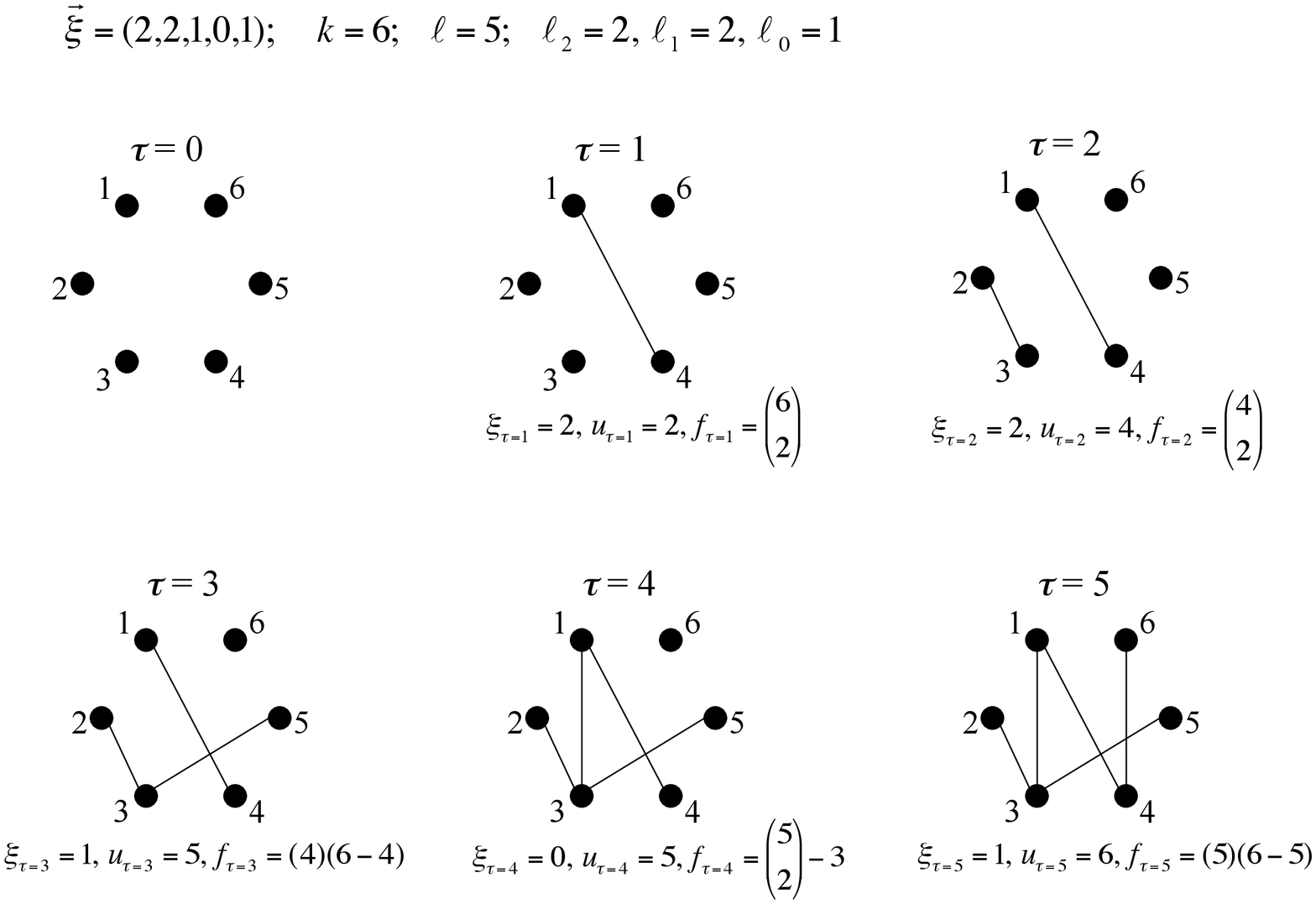,angle=270,scale=0.5}
\caption{Assembly of a conditioned graph contributing to $Q_2(k,\ell)$ for $k=6,\ell=5$. In this example,
the assembly history is given by $\vxi=(2,2,1,0,1)$, representing the fact that
the edges added are, in order of appearance, of types $\ell_2,\ell_2,\ell_1,\ell_0,\ell_1$.
At each assembly step $\tau$, the type of edge $\xi_\tau$, total number of used (discovered) nodes $u_\tau$, and
combinatorial factor $f_\tau$ are given.
The total combinatorics of assemblies with this same history $\vxi=(2,2,1,0,1)$ is given by
Eq.~(\ref{Fxi}).}
\label{assembly-illustration}
\end{figure}

In order to determine $Q_2(k,\ell)$ via assembly, one begins with $k$ isolated nodes and adds
edges, totalling $\ell$, so that every node is connected to an edge. Once two nodes
have been connected by an edge, they cannot be connected again (i.e., multigraphs are not allowed). 
To find all distinct 
graphs that contribute to $Q_2(k,\ell)$, one first needs to determine all possible ways to assemble 
those graphs. The number of distinct assemblies is larger than $Q_2(k,\ell)$, but is trivially corrected to yield
$Q_2(k,\ell)$, as explained below.
For the assembly process, the critical ingredient is knowledge of the number of distinct ways in which a given edge
can enter into the graph. 
I now proceed to describe this enumeration (refer throughout this
section to Fig.~\ref{assembly-illustration} for a specific example of assembly, along with the relevant notation). 

Consider the initial state of $k$ isolated nodes. At this initial step of the process,
there are ${k\choose 2}$ possible pairs of nodes in which an edge can be placed. After the first
step of edge addition, 2 nodes become used (or discovered).
Let us define the vector $\vxi$ which characterizes the edge addition process. This vector has
length $\ell$ (i.e. its dimension $\dim\vxi=\ell$). The $\tau$-th component of the vector, $\xi_\tau$, 
is the number of nodes that are discovered by the addition of the $\tau$-th edge in the assembly;
for the first step, $\xi_{\tau=1}=2$.
Another useful definition is $u_{\tau}$, the number of nodes discovered up to step $\tau$.
In all assemblies, $\xi_{\tau=1}=2$ and $u_{\tau=1}=2$. After the first edge is added
in one of the possible ${k\choose 2}$ places on the graph, there are in total
${k\choose 2}$ possible distinct graphs.

Each additional added edge generates an enumeration depending on the nodes
that are involved in that edge. To illustrate this, consider the possibilities when adding the second edge, 
i.e. $\tau=2$. The first possibility
corresponds to edge $\tau=2$ being used to discover two new nodes out of the remaining $k-2$ undiscovered 
nodes, among which there are ${k-2\choose 2}$ possible node pairs.
This leads to a total of ${k\choose 2}{k-2\choose 2}$ distinct graph assemblies,
where $u_{\tau=2}=4$. In $\vxi$, component $\xi_{\tau=2}=2$ because the second edge discovers 2 new nodes.
The second possibility for $\tau=2$ corresponds to adding an edge that connects 
one of the two nodes already in use to one of the $k-2$ undiscovered nodes.
This leads to ${k\choose 2} (2(k-2))$ distinct graph assemblies because
the second edge has 2 choices among discovered nodes and $k-2$ choices among undiscovered nodes; 
in this case $u_{\tau=2}=3$ and $\xi_{\tau=2}=1$. 

The condition of visiting all $k$ nodes at least once imposes in turn conditions of the numbers
of edges with $\xi_\tau$ equal to 1 or 2.
It is convenient to introduce notation for these edges. If the addition of an edge at a given step $\tau$ 
discovers two unused nodes,
this edge is counted into $\ell_2$ and is described as being a {\it type $\ell_2$ edge}. On the other hand, if 
at $\tau$ an
edge connects a node already discovered in a step $<\tau$ to an undiscovered node, it counts into $\ell_1$ and is
referred to as a {\it type $\ell_1$ edge}. For an arbitrary step $\tau$ in the assembly, type $\ell_2$ edges
are associated with a factor ${k-u_{\tau -1}\choose 2}$ in the enumeration because they connect 2 out of the
remaining $k-u_{\tau -1}$ undiscovered nodes; type $\ell_1$ edges are associated
with a factor $u_{\tau -1}(k-u_{\tau -1})$ because they connect one of the $u_{\tau -1}$ discovered nodes 
to one of the $k-u_{\tau -1}$ undiscovered nodes.
The counts $\ell_1$ and $\ell_2$ are part of the total number
of edges $\ell$. Another kind of edge is possible, which connects two nodes already discovered; these edges
are counted by $\ell_0$ and referred to as {\it type $\ell_0$ edges} 
(the 0 refers to the fact that their introduction
does not contribute to $k$ because they do not discover new nodes). Below, I will give examples of the enumeration
for type $\ell_0$ edges. The relation between $\ell_2,\ell_1,\ell_0$ and $k,\ell$ is summarized in the equations
\begin{eqnarray}
\ell&=&\ell_2+\ell_1+\ell_0
\label{l2-eq}\\
k&=&2\ell_2+\ell_1,
\label{kl2-eq}
\end{eqnarray}
where only integer non-negative solutions are allowed.

At this point, it is useful to make a few simple calculations that illustrate the ideas just described.
First, consider $k$ even, and let us assemble a conditioned graph with the minimum number of edges possible.
Clearly, $\ell=k/2$, where each edge must connect a new pair of undiscovered nodes until all nodes are discovered, 
and therefore $\ell=\ell_2=k/2$.  Hence, there are $\prod_{a=1}^{k/2}{k-2(a-1)\choose 2}$ distinct assemblies and
\begin{equation}
Q_2(k,\ell=k/2)=\frac{1}{\left(\frac{k}{2}\right)!}\prod_{a=1}^{k/2}{k-2(a-1)\choose 2}
=\frac{k!}{2^{k/2}(k/2)!}\qquad
[\text{$k$ even}]
\label{Q2-keven-min}
\end{equation} 
distinct conditioned graphs. The 
$(k/2)!$ in the denominator comes from the fact that the order in which the $\ell=k/2$ edges are chosen
is irrelevant to the assembled graph, and thus their permutations must be taken away. In
$\vxi$ notation, $\vxi=(2,\dots,2)$, i.e., $\xi_{\tau}=2$ for all $\tau$, and $\dim\vxi=\ell_2=k/2$. 
In this example, $\vxi$ is unique.

The next example to consider is when $k$ is an odd number and $\ell$ is minimal 
($\ell=\lceil k/2\rceil=(k+1)/2$ in this case). To assemble such conditioned graphs,
any one of the $k$ nodes must be reused exactly once to achieve the condition of all nodes being
connected to at least one edge, and thus $\ell_1=1$ and $\ell_2=(k-1)/2$.
As before, one chooses the first edge out of ${k\choose 2}$ possibilities, and from $\tau=2$ and beyond
the possibility to add the single edge of type $\ell_1$ is available. If this edge is added in step $\tau=b$, 
and summing over all possible values of $b$, the enumeration becomes
\begin{multline}
Q_2(k,\ell=(k+1)/2)=\\ \frac{1}{\left(\frac{k+1}{2}\right)!}
\sum_{b=2}^{(k+1)/2}(2(b-1))(k-2(b-1))\prod_{a_1=1}^{b-1}{k-2(a_1-1)\choose 2}
\prod_{a_2=b+1}^{(k+1)/2}{k-2(a_2-1)+1\choose2}\\
=\frac{k!}{2^{(k-1)/2}\left(\frac{k-3}{2}\right)!}
\qquad[\text{$k$ odd}],
\label{Q2-kodd-min}
\end{multline}
because for given $b$ there are 
${k\choose 2}{k-2\choose 2}\dots{k-2(b-2)\choose 2}$ ways to assemble the first $2(b-1)$ nodes using type 
$\ell_2$ edges, $(2(b-1))(k-2(b-1))$ possible ways to introduce the type $\ell_1$ edge, and after that, 
there are still $k-2(b-1)-1$ remaining nodes that are connected in ${k-2(b-1)-1\choose 2}\dots{2\choose 2}$ 
possible ways ($k-2(b-1)-1$ is even). The denominator $\left(\frac{k+1}{2}\right)!$ corrects for order in
the permutations of the edge addition. In $\vxi$ notation, each value of $b$ is associated with
a $\vxi$ in which $\xi_{\tau=b}=1$ and $\xi_{\tau\neq b}=2$, and $\dim\vxi=(k+1)/2$. In this
example, unlike before, $\vxi$ is not unique; there are $(k-1)/2$ different $\vxi$, one for each
choice of $b$ between 2 and $(k+1)/2$.

For the last example, consider $k$ even and $\ell=k/2+1$ (no longer minimal). In this situation, one can either: 
i) connect all pairs of nodes by use of $\ell_2=k/2$ edges while at some step $\tau=c$ use a single 
type $\ell_0$ edge to connect two nodes of the $u_{\tau-1=c-1}$ that have already been discovered 
up to step $\tau-1=c-1$, or ii) connect $k-2$ nodes via $\ell_2=(k-2)/2$ edges, and also use $\ell_1=2$ edges at steps
$\tau=b_1$ and $\tau=b_2$ to connect the remaining 2 nodes; the two cases are mutually exclusive.
Therefore, considering all possible $b_1,b_2,c$, 
\begin{multline}
Q_2\left(k,\ell=\frac{k}{2}+1\right)=\frac{1}{\left(\frac{k}{2}+1\right)!}\\
\left[\sum_{b_1=2}^{k/2}(2(b_1-1))(k-2(b_1-1))\sum_{b_2=b_1+1}^{k/2+1}(2(b_2-1)-1)(k-2(b_2-1)+1)\right.\\
\prod_{a_1=1}^{b_1-1}{k-2(a_1-1)\choose 2}\prod_{a_2=b_1+1}^{b_2-1}{k-2(a_2-1)+1\choose2}
\prod_{a_3=b_2+1}^{k/2+1}{k-2(a_3-1)+2\choose 2}\\
\left.+\sum_{c=2}^{k/2}\left({2c\choose 2}-c\right)\prod_{a_1=2}^{k/2}{k-2(a_1-1)\choose 2}\right]
=\frac{k!}{2^{k/2}\left(\frac{k}{2}-2\right)!}\left(\frac{3k+4}{6}\right)
\qquad[\text{$k$ even}].
\label{firstl0}
\end{multline}
The first set of sums in the square brackets enumerate the cases of two separate instances of visiting
one used node ($\ell_2=(k-2)/2,\ell_1=2,\ell_0=0$), and the second sum enumerates the cases when one edge 
is placed between two previously used nodes ($\ell_2=k/2,\ell_1=0,\ell_0=1$). For the second sum, 
note that any type $\ell_0$ edge placed between two nodes already present occurs
when 4 or 6 or ... $k$ nodes have been used for the first time. 
At each of these steps, the number of choices is ${4\choose 2}-2,{6\choose 2}-3,\dots,{k\choose 2}-k/2$,
which account for the number of possible edges between the nodes present minus the edges that have
been placed. Generally, for a type $\ell_0$ edge introduced at step $\tau$, the factor associated with 
its enumeration is ${u_{\tau -1}\choose 2}-(\tau -1)$. Note that for such an edge $u_{\tau}=u_{\tau-1}$.
Once again, the prefactor $1/(1+k/2)!$ accounts for eliminating the permutations among overall edge placement order.
In $\vxi$ notation, there are now two kinds of vectors: for $\ell_2=(k-2)/2,\ell_1=2,\ell_0=0$, there
are ${(k-1)/2 \choose 2}$ distinct $\vxi$, one for each $\xi_{\tau=b_1}=1,\xi_{\tau=b_2}=1,
\xi_{\tau\neq b_1,b_2}=2$; for $\ell_2=k/2,\ell_1=0,\ell_0=1$ there are $(k-2)/2$ vectors $\vxi$,
one for each case of $\xi_{\tau=c}=0,\xi_{\tau\neq c}=2$, where $c\geq 2$ because in this example 
there must be at least 2 edges (and 4 nodes) before any type $\ell_0$ can be introduced.

It is clear that in all examples above, one can use a shorthand to represent the sums for the assemblies by using
$\vxi$. Thus, $Q_2(k,\ell)=(\ell!)^{-1}\sum_{\vxi}C(\vxi)$ where $C(\vxi)$ is the combinatorial factor 
associated with
an assembly history $\vxi$, and the $\vxi$ are chosen to satisfy the given $k$ and $\ell$.
\subsubsection{Setup of the $Q_2(k,\ell)$ calculation}
\label{Q2-strategy}
The three calculations above exhibit all types of edges in the assembly process:
edges that visit two new nodes, edges that visit one new node and a previously visited node, and 
edges that visit two already visited nodes.
Clearly, the kinds and numbers of edges used are constrained to satisfy the 
definition of $Q_2(k,\ell)$ as explained below.
%Enough edges need to be used in order to satisfy the
%condition of each node being connected to an edge.
The function that each edge performs (type $\ell_2,\ell_1$ or $\ell_0$) depends on the step at which it is added,
which is equivalent to assuming that edges are distinguishable. 
The advantage of making this distinguishability available is that it converts the counting of $Q_2(k,\ell)$
into a process 
that is tractable, i.e., it provides rules to count all possibilities.
However, if one looks at the final product of the assembly, the relevant conditioned graphs 
of $Q_2(k,\ell)$, 
it would be impossible to determine which edge came first or what function it performed
(this is the reason why one divides $\sum_{\vxi}C(\vxi)$ by $\ell!$).
Essentially, $Q_2(k,\ell)$ is calculated by first enumerating all possible assemblies
that lead to the conditioned graphs, and then taking away the edge permutations.

As it was shown in Eq.~(\ref{firstl0}), there are multiple choices of $\ell_0,\ell_1,\ell_2$ for given 
$k$ and $\ell$.
Given that in $Q_2(k,\ell)$, $k$ and $\ell$ are specified, it is necessary to express
the conditions on $\ell_0,\ell_1,\ell_2$ as functions of $k$ and $\ell$. 
But one cannot solve for all three $\ell_0,\ell_1,\ell_2$ from Eqs.~(\ref{l2-eq}) and (\ref{kl2-eq}). 
However, it is possible to solve for $\ell_1,\ell_2$ by focusing on $\ell-\ell_0$ and $k$.
The solution is $\ell_2=k-(\ell-\ell_0)$ and $\ell_1=2(\ell-\ell_0)-k$. 
By taking $\ell_0$ as a free parameter, and running over all its possible values,
all triplets $\ell_0,\ell_1,\ell_2$ are uniquely specified. 
All that remains is to determine the allowed range for $\ell_0$ which emerges from
determining the minimum and
maximum $\ell_1 +\ell_2$ ($=\ell-\ell_0$) necessary to visit $k$ nodes, while keeping in mind that $\ell_2\geq 1$
since the first edge is always type $\ell_2$:
the minimum occurs when $\ell_2=\left\lfloor\frac{k}{2}\right\rfloor$ and
$\ell_1=k-2\left\lfloor\frac{k}{2}\right\rfloor$ (which gives $\ell_1=0$ or 1) so 
$\ell_1+\ell_2=\left\lceil\frac{k}{2}\right\rceil$;
the maximum occurs for $\ell_2=1$ and $\ell_1=k-2$ with $\ell_1+\ell_2=k-1$. 
Therefore, $\left\lceil\frac{k}{2}\right\rceil\leq\ell_1+\ell_2\leq k-1$ leading to
$\ell-\left\lceil\frac{k}{2}\right\rceil\geq\ell_0\geq\ell-(k-1)$.
For each unique triplet $\ell_0,\ell_1,\ell_2$, one can define
the number of conditioned graph assemblies $F(\ell_2,\ell_1,\ell_0)$, and 
\begin{equation}
Q_2(k,\ell)=\sum_{\ell_0=\ell-(k-1)}^{\ell-\left\lceil\frac{k}{2}\right\rceil}F(\ell_2,\ell_1,\ell_0)
=\sum_{\ell_0=\ell-(k-1)}^{\ell-\left\lceil\frac{k}{2}\right\rceil}
\sum_{\vxi\in\vxiconf}F(\vxi),
\label{Q2sum}
\end{equation}
where $\vxiconf=\vxiconfshort$ corresponds to the set of all allowed histories 
$\vxi$ consistent with $\ell_0,\ell_1,\ell_2$.
Each $F(\vxi)$ has the form
\begin{equation}
F(\vxi)=(\ell!)^{-1}C(\vxi)
=(\ell!)^{-1}\prod_{\tau=1}^{\ell} f_{\tau}(u_{\tau-1},\xi_{\tau})
\label{Fxi}
\end{equation}
where $f_{\tau}(u_{\tau-1},\xi_{\tau})$ corresponds to the combinatorial assembly factor associated
with the addition of the edge of type $\xi_{\tau}$ at step $\tau$, at which point the 
number of discovered nodes is $u_{\tau-1}$. As stated before, 
$f_{\tau}(u_{\tau-1},\xi_\tau=2)={k-u_{\tau-1}\choose 2}$,
$f_{\tau}(u_{\tau-1},\xi_\tau=1)=u_{u_{\tau-1}}(k-u_{\tau-1})$, and
$f_{\tau}(u_{\tau-1},\xi_\tau=0)={u_{\tau-1}\choose 2}-(\tau-1)$.
The number of used nodes up to step $\tau$ is given by 
\begin{equation}
u_{\tau-1}=\sum_{\tau'=1}^{\tau-1}\xi_{\tau'},
\label{utau}
\end{equation}
which completes the calculation. 

However, given that calculating $Q_2(k,\ell)$ involves
summing over all possible $\vxi$, further specification is possible with more concrete results.
Below, the calculation of $F(\ell_2,\ell_1,\ell_0)$ is tackled in steps by
first addressing $F(\ell_2,\ell_1,\ell_0=0)$ and then using this result to introduce $\ell_0$ edges 
and complete the calculation of $Q_2(k,\ell)$.
\subsubsection{Calculation of $F(\ell_2,\ell_1,\ell_0=0)$}
When $\ell_0=0$, only the combinatorics of $\ell_1$ and $\ell_2$ edges are needed. 
It is useful to introduce the redefinition $(\vxi,\tau)\to(\vec{h},t)$ in this case (the reason becomes
clear in the next Sec.). In this notation $\dim\vec{h}=\ell_2+\ell_1$, and each component $h_t$ can only be 1 or 2. 
The difference between two histories $\vec{h}$ and $\vec{h}'$ with equal $\ell_1$ and $\ell_2$ 
is found in the specific steps $t$ in which $h_t=1$, i.e., the steps in which the type $\ell_1$ 
edges are introduced. It is then convenient to define a set $\{b_1,\dots,b_{\ell_1}\}$ corresponding
to the steps $t$ of the first, second, ..., $\ell_1$ introductions of type $\ell_1$ edges, and 
a counter $\lambda$ from 0 to $\ell_1$.
For a concrete $\vec{h}$, for $\lambda=1$ there is an associated step $t=b_1$. The conditioned graph 
enumeration due to type $\ell_2$ edges up to $t=b_1-1$ is $\prod_{a_1=1}^{b_1-1}{k-2(a_1-1)\choose 2}$
and at $t=b_1$ the new factor $2(b_1-1)(k-2(b_1-1))$ comes in. 
Between the $\lambda-1$ and $\lambda$ edges of type $\ell_1$, that is
steps $b_{\lambda-1}+1\leq t\leq b_{\lambda}-1$, enumeration
due to type $\ell_2$ edges is $\prod_{a_{\lambda}=b_{\lambda-1}+1}^{b_{\lambda}-1}{k-2(a_{\lambda}-1)+(\lambda-1)
\choose 2}$, and at $t=b_{\lambda}$ there is a new factor $(2(b_{\lambda}-1)-(\lambda-1))(k-2(b_{\lambda}-1)+(\lambda-1))$.
These considerations lead to the expression
\begin{multline}
F(\ell_2,\ell_1,\ell_0=0)=\equiv\F0=\frac{1}{(\ell_2+\ell_1)!}\sum_{b_1=2}^{\ell_2+1}[2(b_1-1)][k-2(b_1-1)]
\prod_{a_1=1}^{b_1-1} {k-2(a_1-1)\choose 2}\dots\\
\sum_{b_{\lambda}=t_{\lambda-1}+1}^{\ell_2+\lambda}[2(b_{\lambda}-1)-(\lambda-1)]
[k-2(b_{\lambda}-1)+(\lambda-1)]\prod_{a_{\lambda}=b_{\lambda-1}+1}^{b_{\lambda}-1}
{k-2(a_{\lambda}-1)+(\lambda-1)\choose 2} \dots\\
\sum_{b_{\ell_1}=b_{\ell_1-1}+1}^{\ell_2+\ell_1}[2(b_{\ell_1}-1)-(\ell_1-1)][k-2(b_{\ell_1}-1)+(\ell_1-1)]
\prod_{a_{\ell_1}=b_{\ell_1-1}+1}^{b_{\ell_1}-1} {k-2(a_{\ell_1}-1)+(\ell_1-1)\choose 2}\\
\prod_{a_{\ell_1+1}=b_{\ell_1}+1}^{\ell_2+\ell_1} {k-2(a_{\ell_1+1}-1)+\ell_1\choose 2},
\label{F0}
\end{multline}
where the sums in Eq.~(\ref{F0}) reflect all possible ways to choose the set of $b_\lambda$.

Equation (\ref{F0}) can be evaluated by noting that the factors due to $\ell_2$ edges together
with the factors of form $[k-2(b_{\lambda}-1)+(\lambda-1)]$ within the type $\ell_1$ enumeration 
combine to the factorial $k!=(2\ell_2+\ell_1)!$.
The denominators coming from the $\ell_2$ factors produce $2^{\ell_2}$. What remains is the
sum of products of the form $[2(b_\lambda-1)-(\lambda-1)]$ which come from type $\ell_1$ edges, and counts the
ways to pick nodes from those that have been discovered in steps previous to $t=b_\lambda$, for all
$\lambda\leq\ell_1$.
Hence, one can write
\begin{equation}
\F0=\frac{k!}{2^{\ell_2}(\ell_2+\ell_1)!}A_1(\ell_2,\ell_1)
\label{F0final}
\end{equation}
where
\begin{equation}
A_1(\ell_2,\ell_1)\equiv \sum_{b_1=2}^{\ell_2+1}\sum_{b_2=b_1+1}^{\ell_2+2}
\dots\sum_{b_{\ell_1}=b_{\ell_1-1}+1}^{\ell_2+\ell_1}[2(b_1-1)][2(b_2-1)-1]\dots[2(b_{\ell_1}-1)-(\ell_1-1)].
\label{A1}
\end{equation}
Equation~(\ref{F0final}) states that the number of ways to assemble the $k$ nodes when $\ell_0=0$ is
proportional to the permutations of the nodes and the number of choices in which single previously used nodes
can be picked (as $\ell_1$ edges are introduced).
In $\vec{h}$ notation, $A_1(\ell_2,\ell_1)$ can be written as
\begin{equation}
A_1(\ell_2,\ell_1)=\sum_{\vec{h}\in\{\vec{h}(\ell_2,\ell_1)\}_{\rm conf}}A_1(\vec{h})
=\sum_{\vec{h}\in\{\vec{h}(\ell_2,\ell_1)\}_{\rm conf}}\prod_{b\in\{t|h_{t}=1\}}u_b,
\end{equation}
where (with a slight abuse of notation) $b$ is an element of $\{t|h_{t}=1\}$, the set of all steps 
in assembly $\vec{h}$ at which a type $\ell_1$ edge is added.
In $\vec{h}$ notation, 
\begin{equation}
F^{(0)}(\ell_2,\ell_1)=\sum_{\vec{h}\in\{\vec{h}\}_{\rm conf}}F^{(0)}(\vec{h})
=\frac{k!}{2^{\ell_2}(\ell_2+\ell_1)!}\sum_{\vec{h}\in\{\vec{h}\}_{\rm conf}}A_1(\vec{h}).
\label{F0s}
\end{equation}
To develop some intuition about $F^{(0)}(\ell_2,\ell_1)$, it is useful to make reference to a few examples: if 
$\ell_2=1$ and thus $\ell_1=k-2$, $F^{(0)}(\ell_2,\ell_1)$ is the number of distinct realizations of invasion
percolation without trapping, where the initial seed is an edge (of indistinguishable nodes). 
For $\ell_2>1$, $F^{(0)}(\ell_2,\ell_1)$
counts a forest of $\ell_2$ of these invasion percolation trees (the trees never coalesce).
\subsubsection{Introducing $\ell_0>0$ and full $Q_2(k,\ell)$}
\label{Q2-calculation}
To introduce an edge of type $\ell_0$, there must be nodes already used and, in addition, pairs of them 
that have not been directly connected by another edge. These unconnected node pairs are {\it vacancies}. 
The combinatorics
of type $\ell_0$ edges require counting the vacancies available as the conditioned graph assembly progresses. 
The availability of vacancies is restricted by the assembly sequence $\vec{h}$. 
For instance, consider the first two steps of any assembly. After the first edge of type
$\ell_2$, the second edge can only be type $\ell_1$ or $\ell_2$, but not type $\ell_0$ because there are no 
vacancies in the graph yet. Using the notation for steps applied when $\ell_0$, the first step at which a
type $\ell_0$ edge can be introduced is right before $t=3$ since there would be 
four vacancies if the second edge is type $\ell_2$ or one vacancy if it is type $\ell_1$ (the distinction
between $\vec{h},t$ and $\vxi,\tau$ becomes more evident in this section, where $t$ can be used to describe 
the equations for the full assembly including type $\ell_0$ edges, even though $t$ only counts steps
that add nodes, whereas $\tau$ counts every edge addition).

Edges of type $\ell_0$ can be placed in any step $t$ of the sequence $\vec{h}$ 
where there are available vacancies, and 
to obtain the full enumeration, all possible placings must be counted. Fortunately, even though placing a type 
$\ell_0$ edge is conditional on the vacancies created by $\ell_2$ and $\ell_1$, the opposite is not true,
i.e., placings of $\ell_1$ and $\ell_2$ are unaffected by $\ell_0$, and thus the results of $\F0$
can be used here.
This is because the combinatorics of type $\ell_1,\ell_2$ edges only
depend on the numbers of used and unused nodes, and type $\ell_0$ edges have no effect on those. 

Following the previous description, it makes sense to introduce 
$\vec{v}=(v_1,v_2,\dots,v_{\ell_1+\ell_2})$,
the vacancies available due to the addition of type $\ell_1$ and $\ell_2$ edges, at the respective steps 
$t=1,2,\dots,\ell_1+\ell_2$ of $\vec{h}$ (clearly $\vec{v}$ is a function of $\vec{h}$). These are the vacancies
where type $\ell_0$ edges can be placed. The values of $v_t$ are defined such that they are 
not affected by the addition of type $\ell_0$ edges. To track type $\ell_0$ edges, one defines
$\vec{n}=(n_1,n_2,n_3,\dots,n_{\ell_1+\ell_2})$, the number of edges type $\ell_0$ 
placed, respectively, {\it immediately after} $t=1,2,3,\dots,\ell_1+\ell_2$ edges of type $\ell_1$ 
and $\ell_2$ have been added (to be clear, at step $t$, an edge of type $\ell_2$ or $\ell_1$ is added, leading
to $v_t$, and before the next step $t+1$, $n_t$ edges of type $\ell_0$ are added).
Both $v_1$ and $n_1$ are equal to zero because there are no vacancies created with the first edge addition
and thus it is valid to omit them from $\vec{n}$ and $\vec{v}$ if desired.
To determine the combinatorial weight of any particular sequence of $\ell_0$ placings, edges can choose
among the available vacancies: at step $t=2$, there are $v_2$ vacancies, and so $0\leq n_2\leq v_2$, 
which can be done in $\frac{v_2!}{(v_2-n_2)!}$ ways (keeping in mind the edges are considered distinguishable
while being assembled); 
at $t=3$, there are $v_3-n_2$ vacancies, and $0\leq n_3\leq v_3-n_2$, with combinatorial 
weight $(v_3-n_2)!/(v_3-n_2-n_3)!$; etc.
Therefore, the number of combinations for the sequences $\vec{v}$ and $\vec{n}$ are
\begin{equation}
A_0(\vec{n},\vec{v}(\vec{h}))=\prod_{t=2}^{\ell_1+\ell_2}
\frac{\left(v_t-\sum_{t'=2}^{t-1} n_{t'}\right)!}{\left(v_t-\sum_{t'=2}^{t} n_{t'}\right)!};
\qquad \left[\text{constrained to }\ell_0=\sum_{t=2}^{\ell_1+\ell_2}n_t\right].
\label{A0nv}
\end{equation}
For a given sequence $\vec{v}$, all allowed $\vec{n}$ contribute to $Q_2(k,\ell)$, and therefore
it is necessary to sum over all $\vec{n}$ subject to the condition in the brackets.
Thus, to each term $A_1(\vec{h})$, one multiplies the factor
\begin{equation}
\sum_{\vec{n}}A_0(\vec{n},\vec{v}(\vec{h}))
=\sum_{[n_2+\dots+n_{\ell_1+\ell_2}=\ell_0]}\prod_{t=2}^{\ell_1+\ell_2}
\frac{\left(v_t-\sum_{t'=2}^{t-1} n_{t'}\right)!}{\left(v_t-\sum_{t'=2}^{t} n_{t'}\right)!},
\label{SumA0nv}
\end{equation}
where the notation of the sum implies summing over all combinations of $n_t$ that satisfy the constraint
$n_2+\dots+n_{\ell_1+\ell_2}=\ell_0$.
To fully specify the previous, and recalling Eq.~(\ref{utau}), $v_t$ is given by
\begin{equation}
v_t={u_t\choose 2}-t,
\label{vh}
\end{equation}
which has already been mentioned in the discussions of Eqs.~(\ref{firstl0}) and (\ref{Fxi}).

These results can now be put together in a single expression. From Eqns.~(\ref{F0s})
and (\ref{SumA0nv})
\begin{equation}
F(\ell_2,\ell_1,\ell_0)=\frac{k!}{2^{\ell_2}(\ell_2+\ell_1+\ell_0)!}
\sum_{\vec{h}}A_1(\vec{h})\sum_{\vec{n}}A_0(\vec{n},\vec{v}(\vec{h})).
\end{equation}
With the use of Eqn.~(\ref{Q2sum}) and the relations between $k,\ell$ and $\ell_2,\ell_1,\ell_0$, 
this translates into the final result
\begin{multline}
Q_2(k,\ell)=\sum_{\ell_0=\ell-(k-1)}^{\ell-\left\lceil\frac{k}{2}\right\rceil}
F(k-(\ell-\ell_0),2(\ell-\ell_0)-k,\ell_0)\\
=\frac{k!}{2^{(k-\ell)}\ell!}
\sum_{\ell_0=\ell-(k-1)}^{\ell-\left\lceil\frac{k}{2}\right\rceil}
\frac{1}{2^{\ell_0}} \sum_{\vec{h}}A_1(\vec{h})\sum_{\vec{n}}A_0(\vec{n},\vec{v}(\vec{h})).
\label{Q2final}
\end{multline}
It is interesting to write down a few results for $Q_2(k,\ell)$ to gain some concrete
intuition of how the numbers evolve as $k$ and $\ell$ change (see Table~\ref{Q2table}).
Evidently, since the sums over $\vec{h}$ and $\vec{n}$ span all possible cases, the effect of specific assembly
histories is summed away, and it is sensible to define a combinatorial coefficient dependent
only on $k,\ell,\ell_0$. Thus
\begin{multline}
A(\ell_2,\ell_1,\ell_0)=A(k-(\ell-\ell_0),2(\ell-\ell_0)-k,\ell_0)
\equiv \sum_{\vec{h}}A_1(\vec{h})\sum_{\vec{n}}A_0(\vec{n},\vec{v}(\vec{h}))\\
= \sum_{t_1=2}^{\ell_2+1}\dots\sum_{t_{\ell_1}=t_{\ell_1-1}+1}^{\ell_2+\ell_1}
\sum_{[n_2+\dots+n_{\ell_1+\ell_2}=\ell_0]}
\prod_{l=1}^{\ell_1}[2(t_l-1)-(l-1)]
\prod_{t=2}^{\ell_1+\ell_2}
\frac{\left(v_t-\sum_{t'=2}^{t-1} n_{t'}\right)!}{\left(v_t-\sum_{t'=2}^{t} n_{t'}\right)!},
\end{multline}
where $v_t$ is defined through Eqns.~(\ref{utau}) and (\ref{vh}). The author is not aware of any
combinatorial identity that allows the previous expression to be reduced further. Clearly,
using the inclusion-exclusion principle, the left and right hand sides of Eq.~(\ref{Q2final})
could be evaluated to write an alternating series for $A$, but this would defeat the
purpose of having only additive terms. Multivariate asymptotics of the expressions inside
the sums are in principle possible in the field of enumerative asymptotics~\cite{Flajolet,Odlyzko,Pemantle}
but techniques are not well suited yet for arbitrary dimension calculations in cases such as $A$.

A straightforward characterization of $Q_2(k,\ell)$ is found in Fig.~\ref{Q2kl-asymptotic}, where the 
plots show $\ln Q_2(k,\ell)$ and $\ln (Q_2(k,\ell)/{{k\choose 2}\choose\ell})$ as functions of $k$ and $\ell$.
It is clear that to a large extent, $Q_2(k,\ell)\to {{k\choose 2}\choose\ell}$ for large enough $\ell$ with 
respect to $k$, but this behavior breaks down when $\ell\sim\left\lceil\frac{k}{2}\right\rceil$. This
limit behavior is also valid for general $r$. Results
for $Q_2(k,\left\lceil\frac{k}{2}\right\rceil)$ (and for general $r$ as well),
where $\ell$ is at its minimum, are presented in Sec.~\ref{Qrmin}.
A full treatment of the asymptotics of $Q_2(k,\ell)$ is presented in Ref.~\cite{Bender-Canfield-McKay,Korshunov}, and 
therefore will not be tackled here.
\subsection{Extension to $Q_{r-1}(k,\ell)$}
\label{Qr-extension}
The treatment above can be extended to arbitrary $r$. A conditioned hypergraph with $\ell$ hyperedges,
of uniform rank $r-1$, where all $k$ nodes are visited by at least one hyperedge, 
can be assembled via hyperedges that are differentiated in terms of the number of visited nodes. 
Each hyperedge can find $0,1,2,\dots, r-1$ nodes as it is placed, leading to the edge types
counted by $\ell_0,\ell_1,\dots,\ell_{r-1}$. The inputs $k$ and $\ell$ satisfy 
\begin{eqnarray}
\ell&=&\ell_0+\ell_1+\ell_2+\dots+\ell_{r-1}
\label{lr-eq}\\
k&=&\ell_1+2\ell_2+\dots+(r-1)\ell_{r-1}.
\label{klr-eq}
\end{eqnarray}
As explained for the case of $Q_2(k,\ell)$ in Sec.~\ref{Q2-strategy}, (virtually) all possible non-negative
integer solutions to 
the Eqns.~(\ref{lr-eq}) and (\ref{klr-eq}) need to be used in order to enumerate all possible 
conditioned hypergraphs that contribute to $Q_{r-1}(k,\ell)$. In the present case, it is less straightforward 
to determine the number of solutions to Eqns.~(\ref{lr-eq}) and (\ref{klr-eq}) than in the $r=3$ case. 
However, it only requires calling upon the definition of integer partitions to give an answer. 

Recall that integer Eq.~(\ref{klr-eq}) on its own 
is in fact the condition satisfied by integer partitions of $k$ in which
the largest part is at most $r-1$~\cite{Riordan,Flajolet}. The number of integer partitions of $x$
with maximum part $y$ ($x,y$ both integers), expressed here as $\wp(x,y)$, has been well studied, 
and is known to satisfy certain 
asymptotic formulas and recurrence relations. To use this definition in the present case, a few details
need to be dealt with because aside from Eq.~(\ref{klr-eq}), both Eq.~(\ref{lr-eq}) and $\ell_{r-1}\geq 1$
(first edge is always type $\ell_{r-1}$) also need to satisfied.
First, one can reduce Eq.~(\ref{lr-eq}) by subtracting $\ell_0$
from $\ell$ because the former hyperedge type has no effect on $k$. Then, eliminating $\ell_1$ between
$\ell-\ell_0$ and $k$ yields $k-(\ell-\ell_0)=\ell_2+2\ell_3+\dots+(r-2)\ell_{r-1}$. In this form, almost
all restrictions have been absorbed, except for $\ell_{r-1}\geq 1$. Making the
change of variables $\ell_{r-1}'\equiv \ell_{r-1}-1$, one can finally write the relation
\begin{equation}
k-(\ell-\ell_0)-(r-2)=\ell_2+2\ell_3+\dots+(r-2)\ell_{r-1}'.
\label{k_l_r_condition}
\end{equation}
Now the variables
$\ell_2,\dots,\ell_{r-2},\ell_{r-1}'$ only need to be non-negative integers. 
Therefore, the number of solutions is equal to $\wp(k-(\ell-\ell_0)-(r-2),r-2)$, 
as by Eq.~(\ref{k_l_r_condition}) $k-(\ell-\ell_0)-(r-2)$ can be partitioned
by any valid combination of $\ell_2$ times 1, $\ell_3$ times 2, ..., $\ell_{r-1}'$ times $r-2$.
Note that for $r=3$, one obtains $\wp(k-(\ell-\ell_0)-1,1)=1$, i.e., the solutions are unique for given
$k,\ell,\ell_0$. 
For arbitrary $r$, the number of values for $\ell_0$ is determined from the limits of
$\ell-\ell_0=\ell_1+\ell_2+\dots+\ell_{r-1}$. The smallest value, called $(\ell_1+\dots+\ell_{r-1})_{\rm min}$ 
occurs when $\ell_{r-1}=\lfloor k/(r-1)\rfloor$ and there is a single additional hyperedge of type 
$\ell_{m}$, where $m=k-(r-1)\lfloor k/(r-1)\rfloor$ (if $m=0$ then exactly $\lfloor k/(r-1)\rfloor$ 
hyperedges are needed); altogether, $(\ell_1+\dots+\ell_{r-1})_{\rm min}=\lceil k/(r-1)\rceil$.
On the other hand, $(\ell_1+\dots+\ell_{r-1})_{\rm max}=k-(r-2)$ due to $\ell_{r-1}=1$ and all other hyperedges 
finding one node at a time, i.e. $\ell_1=k-(r-1)$. Therefore, 
$\lceil k/(r-1)\rceil\leq\ell-\ell_0\leq 1+k-(r-1)$ which means
$\ell-(k-r+2)\leq\ell_0\leq\ell-\lceil k/(r-1)\rceil$.
With these considerations, the number of solutions to Eqns.~(\ref{lr-eq}) and (\ref{klr-eq}) is 
\begin{multline}
\sum_{\ell_0=\ell-(k-r+2)}^{\ell-\lfloor k/(r-1)\rfloor} \wp(k-(\ell-\ell_0)-(r-2),r-2)\\
=\sum_{\ell_0=\ell-(k-r+2)}^{\ell-(k-2(r-2))} \wp(k-(\ell-\ell_0)-(r-2))
+\sum_{\ell_0=\ell-(k-2(r-2))+1}^{\ell-\lfloor k/(r-1)\rfloor} \wp(k-(\ell-\ell_0)-(r-2),r-2)
\end{multline}
where $\wp(k-(\ell-\ell_0)-(r-2))$ is the number of integer partitions with no restriction. 
The second sum in the last equality occurs because the restriction of the largest number to be $r-2$
begins to apply for $\ell_0\geq\ell-k+2(r-2)+1$; if $k\leq 2(r-1)$ this term drops out. For small
$r$ such as 3,4,5, these expressions can be studied exactly, by obtaining expressions
for restricted $\wp(x,y)$ from recurrence relations, and maybe using tables for unrestricted $\wp(x)$.
For instance, with the recurrence relation $\wp(x,y)=\wp(x,y-1)+\wp(x-y,y)$ and boundary conditions
$\wp(x,0)=0$, $\wp(1,y)=1$, and $\wp(x,y\geq x)=\wp(x)$~\cite{Andrews}, one obtains
$\wp(x,1)=1$ and $\wp(x,2)=\lceil x/2\rceil$. As $r$ increases, asymptotics become necessary. 
Classic results are available in this area such as the Hardy-Ramanujan asymptotics 
$\wp(x)\sim \exp(\pi\sqrt{2x/3})/(4x\sqrt{3})$ and the asymptotics of
restricted partitions $\wp(x,y)\sim x^{y-1}/[y!(y-1)!]$~\cite{Flajolet}. 

To complete this section, I describe the combinatorics of the placing of hyperedges in the assembly process
that leads to $Q_{r-1}(k,\ell)$. 
In the general case, a hyperedge of type $\ell_m$ (with $1\leq m\leq r-1$) chooses $m$ 
unused nodes and $r-1-m$ used nodes. At any given step $\tau$ of the assembly, there are $u_{\tau-1}$ nodes 
that have been used, and $k-u_{\tau-1}$ that are yet to be used. The hyperedge at step $\tau$ has a combinatorial
factor ${u_{\tau-1}\choose r-1-m}{k-u_{\tau-1}\choose m}$. 
Type $\ell_0$ hyperedges 
are added in the vacancies that other hyperedges provide, and their combinatorics
are no different qualitatively than in the case $r=3$: for $u_{\tau-1}$ used nodes, there are 
${u_{\tau-1}\choose r-1}-(\tau-1)$ vacancies.
The combinatorial contribution of each assembly history $\vxi$ is given by Eq.~(\ref{Fxi}) with
\begin{equation}
f_\tau(u_{\tau-1},\xi_\tau)=\left\{
\begin{array}{lc}
{u_{\tau-1}\choose r-1-\xi_\tau}{k-u_{\tau-1}\choose \xi_\tau};&\qquad [1\leq \xi_{\tau}\leq r-1]\\
{u_{\tau-1}\choose r-1}-(\tau-1);&\qquad[\xi_\tau=0],
\end{array}
\right.
\label{xi-r-rules}
\end{equation}
and $u_\tau=\sum_{\tau'=1}^\tau\xi_\tau$.
Although it is possible to write down the expression for $Q_{r-1}(k,\ell)$, its cumbersome nature would 
not add much new intuition. However, the combinatorial rules in Eq.~(\ref{xi-r-rules}) are used in Sec.~\ref{Qrmin} 
to calculate $Q_{r-1}(k,\ell)$ when $\ell$ is at its minimum value $\lceil k/(r-1)\rceil$.
\section{Useful results concerning $Q_{r-1}(k,\ell)$}
\label{identities-sec}
\subsection{Some identities of $Q_{r-1}(k,\ell)$, normalization of $\psi_i(k_i,p)$, and moments 
$\langle k^q_i\rangle$}
\label{identities}
The calculation of $\langle k_i\rangle$ for arbitrary $r$ boils down to
\begin{multline}
\langle k_i\rangle=\sum_{k_i=0}^{N-1}k_i\psi_i(k_i,p)\\
=\sum_{k_i=0}^{N-1}k_i{N-1\choose k_i}\sum_{\ell_i=\lceil k_i/(r-1)\rceil}^{{k_i\choose r-1}}
Q_{r-1}(k_i,\ell_i)p^{\ell_i}(1-p)^{{N-1\choose r-1}-\ell_i}\\
=\sum_{\ell_i=0}^{{N-1\choose r-1}}p^{\ell_i}(1-p)^{{N-1\choose r-1}-\ell_i}
\sum_{k_i}k_i{N-1\choose k_i}Q_{r-1}(k_i,\ell_i),
\end{multline}
This calculation requires solving the sum $\sum_{k_i}k_i{N-1\choose k_i}Q_{r-1}(k_i,\ell_i)$ for all allowed
values of $k_i$. This evaluation can be done by reinserting the inclusion-exclusion 
expression for $Q_{r-1}$ and using a generating function approach on the key sum.
To be specific, 
\begin{multline}
\sum_{k_i}k_i{N-1\choose k_i}Q_{r-1}(k_i,\ell_i)=\sum_{k_i}k_i{N-1\choose k_i}
\sum_{m=0}^{k_i}(-1)^{k_i-m}{k_i\choose m}{{m\choose r-1}\choose \ell_i}\\
=\sum_{m}(-1)^{-m} {{m\choose r-1}\choose \ell_i}
\sum_{k_i}(-1)^{k_i}k_i{k_i\choose m}{N-1\choose k_i},
\end{multline}
where again $i$ is dropped when appropriate because it is irrelevant for these identities.
One can then show, using generating functions (below), that
\begin{multline}
\sum_{k}(-1)^{k}k{k\choose m}{N-1\choose k}\\
=(-1)^{N-1}(N-1)\left[{N-2\choose m}\delta_{m,N-2}+{N-1\choose m}\delta_{m,N-1}\right]
=(-1)^{N-1}(N-1)\left[\delta_{m,N-2}+\delta_{m,N-1}\right], 
\label{Q-comb-ident}
\end{multline}
leading to
\begin{multline}
\sum_{k_i}k_i{N-1\choose k_i}Q_{r-1}(k_i,\ell_i)=\sum_{k_i}k_i{N-1\choose k_i}
\sum_{m=0}^{k_i}(-1)^{k_i-m}{k_i\choose m}{{m\choose r-1}\choose \ell_i}\\
=\sum_{m}(-1)^{-m} {{m\choose r-1}\choose \ell_i}(-1)^{N-1}(N-1)
\left[{N-2\choose m}\delta_{m,N-2}+{N-1\choose m}\delta_{m,N-1}\right]\\
=(N-1){{N-1\choose r-1}\choose \ell_i}-(N-1){{N-2\choose r-1}\choose \ell_i}.
\label{ki_Q_sum}
\end{multline}
Therefore, $\langle k_i\rangle$ becomes
\begin{multline}
\langle k_i\rangle=\sum_{\ell_i=0}^{{N-1\choose r-1}}p^{\ell_i}(1-p)^{{N-1\choose r-1}-\ell_i}
\left[(N-1){{N-1\choose r-1}\choose \ell_i}-(N-1){{N-2\choose r-1}\choose \ell_i}\right]\\
=(N-1)\left[1-(1-p)^{{N-2\choose r-2}}\right],
\end{multline}
where ${N-1\choose r-1}={N-2\choose r-1}+{N-2\choose r-2}$ has been used in 
$(1-p)^{{N-1\choose r-1}-\ell_i}=(1-p)^{{N-2\choose r-1}+{N-2\choose r-2}-\ell_i}$.

To show Eq.~(\ref{Q-comb-ident}), let us define
\begin{equation}
b_{N-1,m}\equiv\sum_{k}(-1)^{k}k{k\choose m}{N-1\choose k}
\end{equation}
and associate to it the generating function~\cite{Riordan,Flajolet,Wilf}
\begin{multline}
b_{N-1}(z)=\sum_m z^m b_{N-1,m}=\sum_m z^m \sum_{k}(-1)^{k}k{k\choose m}{N-1\choose k}\\
=\sum_{k}(-1)^{k}k{N-1\choose k}\sum_m {k\choose m}z^m\\
=\sum_{k}(-1)^{k}k{N-1\choose k}(1+z)^{k}.
\end{multline}
Noting that 
\begin{equation}
(1+z)\frac{d}{dz}\sum_{k}(-1)^{k}{N-1\choose k}(1+z)^{k}
=\sum_{k}(-1)^{k}k{N-1\choose k}(1+z)^{k},
\end{equation}
and using
\begin{equation}
\sum_{k}(-1)^{k}{N-1\choose k}(1+z)^{k}=(-z)^{N-1},
\end{equation}
one obtains
\begin{equation}
b_{N-1}(z)=\sum_{k}(-1)^{k}k{N-1\choose k}(1+z)^{k}=(1+z)\frac{d}{dz}(-z)^{N-1}
=(N-1)(-1)^{N-1}(1+z)z^{N-2}.
\end{equation}
To obtain the $m$-th coefficient of $b_{N-1}(z)$, one can apply
\begin{multline}
b_{N-1,m}=\frac{1}{m!}\left.\frac{d^m}{dz^m}b_{N-1}(z)\right|_{z=0}
=(-1)^{N-1}(N-1)\left[{N-2\choose m}\delta_{m,N-2}+{N-1\choose m}\delta_{m,N-1}\right]\\
=(-1)^{N-1}(N-1)\left[\delta_{m,N-2}+\delta_{m,N-1}\right]
\end{multline}
confirming Eq.~(\ref{Q-comb-ident}).

Higher moments $\langle k^q_i\rangle$ can be calculated through a generalization of the previous result, namely
\begin{equation}
\sum_{k}(-1)^k k^q{k\choose m}{N-1\choose k}=\frac{(-1)^{N-1}}{m!}\frac{d^m}{dz^m}\left.\left[
\left(\frac{d}{dz}+z\frac{d}{dz}\right)^qz^{N-1}\right]\right|_{z=0},
\label{identity_kq}
\end{equation}
where the parenthesis to the power $q$ is to be looked at as an operator that needs to be 
expanded for specific $q$. For instance, for $q=2$, this identity leads to
\begin{equation}
\langle k_i^2\rangle=(N-1)^2-(N-1)(2N-3)(1-p)^{N-2\choose r-2}+(N-1)(N-2)(1-p)^{{N-2\choose r-2}+{N-3\choose r-2}}.
\end{equation}

The normalization of $\psi_i(k_i,p)$ can be confirmed by using 
\begin{equation}
{{N-1\choose r-1}\choose \ell}=\sum_{k}{N-1\choose k}Q_{r-1}(k,\ell)
\label{Q_sum_inverse}
\end{equation}
which simply states that the number of ways in which to choose $\ell$ distinct hyperedges of rank 
$r-1$ out of a total of ${N-1\choose r-1}$ possibilities is equal to the sum of
taking $k$ elements out of $N-1$, weighted by the number of ways in which those $k$ elements
form $\ell$ groups of size $r-1$ such that no element goes unused ($Q_{r-1}(k,\ell)$). The expression
can be shown algebraically via generating functions, in the same kind of approach as above. Also,
it can be obtained by direct application of Eq.~(\ref{identity_kq}) with $q=0$.
\subsection{$Q_{r-1}(k,\ell=\ell)$ for minimum $\ell=\lceil k/(r-1)\rceil$}
\label{Qrmin}
Given that in the sparse regime $\psi_i(k_i,p)$ is dominated by the contribution of the minimum
number of hyperedges $\ell_i=\lceil k_i/(r-1)\rceil$ necessary to visit $k_i$ neighbors, 
Eq.~(\ref{psi-poisson}) requires calculating $Q_{r-1}(k_i,\lceil k_i/(r-1)\rceil)$. 
The case for $r=3$ was derived in Eqns.~(\ref{Q2-keven-min}) and (\ref{Q2-kodd-min}), giving
\begin{equation}
Q_2\left(k,\left\lceil\frac{k}{2}\right\rceil \right)
=\left\{
\begin{array}{ll}
\frac{k!}{2^{k/2}\left(\frac{k}{2}\right)!}&\qquad; k\text{ even}\\
\frac{k!}{2^{(k-1)/2}\left(\frac{k-3}{2}\right)!}&\qquad; k\text{ odd}.
\end{array}
\right.
\end{equation}

Extending this result to general $r$ is straightforward for the case when $k$ is an
exact multiple of $r-1$, so that $k=j(r-1)$ with $j$ an integer. In this case, each
node is part of a single $r-1$-clique, and no cliques overlap. The number $j$
is the exact number of cliques needed to visit the $k$ nodes. The first $r-1$ 
nodes are chosen from $k$ in ${k\choose r-1}$ ways, the next nodes are chosen in
${k-(r-1)\choose r-1}$ ways, etc. After $j$ steps, and recalling the need to compensate
for the permutation of hyperedges (or cliques), one arrives at
\begin{multline}
Q_{r-1}\left(k,\left\lceil\frac{k}{r-1}\right\rceil \right)
=\frac{1}{\left(\frac{k}{r-1}\right)!}{k\choose r-1}{k-(r-1)\choose r-1}
\dots{r-1\choose r-1}\\
=\frac{1}{\left(\frac{k}{r-1}\right)!}
\frac{k!}{\left[(r-1)!\right]^{k/(r-1)}}\qquad [k/(r-1)\text{ positive integer}].
\end{multline}

The more complicated case emerges when $k=j(r-1)+g$, where both $j$ and $g$ are positive 
integers and $1\leq g\leq r-2$, because it means that the $k$ nodes have to be visited
by a total of $\ell=j+1$ hyperedges ($\ell$ is minimum since $j+1=\lceil k/(r-1)\rceil$). This, however, allows 
considerable freedom. Let us enumerate the $j+1$ steps involved in visiting the $k$ nodes
by the index $t$. For $t=1$, exactly $r-1$ nodes are visited. For $t=2$, the second hyperedge
can visit in principle any number of new nodes between 1 and $r-1$. Let us define $d_t$ as the difference between 
$r-1$ and the number of new nodes visited in step $t$. Note that $d_1=0$ by definition.
After $t=j$ steps have occurred, one finds
\begin{equation}
\sum_{t=1}^j d_t=\sum_{t=2}^j d_t=d.
\end{equation}
At $t=j$, there are $g+d$ unvisited nodes which must satisfy $g\leq d+g\leq r-1$ (so the last
hyperedge can visit the remaining unvisited nodes), leading to $0\leq d\leq r-1-g$.
To make use of these facts, one must first calculate the combinatorial
weight of a specific set of values for $d_t$, and then sum over all the choices.
The calculation hinges on determining the combinatorial weight of a single step $t$. 
At this step, $(t-1)(r-1)-\sum_{t'=2}^{t-1}d_{t'}$ nodes have already been visited, 
$k-(t-1)(r-1)+\sum_{t'=2}^{t-1}d_{t'}$ remain unvisited, and the $t$-th
hyperedge visits $r-1-d_t$ new nodes. This leads to the combinatorial factor
\begin{equation}
f_t= {k-(t-1)(r-1)+\sum_{t'=2}^{t-1}d_{t'}\choose (r-1)-d_t} {(t-1)(r-1)-\sum_{t'=2}^{t-1}d_{t'}\choose d_t}.
\end{equation}
The first of the two binomials counts the choices in picking unused nodes, and the 
second counts the choices of previously used nodes. After $j$ steps, the unused nodes equal $g+d$,
and the used nodes are $k-g-d$, and the last hyperedge must pick $r-1-g-d$ from the later. 
Therefore, for given set $\{d_t\}_{1\leq t\leq j}$, the total number of choices is
\begin{equation}
{k-g-d\choose r-1-g-d}{g+d\choose g+d}
\prod_{t=1}^j {k-(t-1)(r-1)+\sum_{t'=2}^{t-1}d_{t'}\choose (r-1)-d_t} 
\prod_{t=1}^j {(t-1)(r-1)-\sum_{t'=2}^{t-1}d_{t'}\choose d_t}.
\end{equation} 
Since $0\leq d\leq r-1-g$, with $g=k-(r-1)\left\lfloor\frac{k}{r-1}\right\rfloor=\text{mod}(k,r-1)$,
and dividing by the permutations over edges, the total 
number of choices becomes
\begin{multline}
Q_{r-1}\left(k,\left\lceil\frac{k}{r-1}\right\rceil \right)
=\frac{1}{\left(\left\lceil\frac{k}{r-1}\right\rceil \right)!}
\sum_{d=0}^{r-1-g}\sum_{[d_2+\dots+d_j=d]}
{k-g-d\choose r-1-g-d}{g+d\choose g+d}\\
\prod_{t=1}^{\left\lfloor\frac{k}{r-1}\right\rfloor} {k-(t-1)(r-1)+\sum_{t'=2}^{t-1}d_{t'}\choose (r-1)-d_t} 
{(t-1)(r-1)-\sum_{t'=2}^{t-1}d_{t'}\choose d_t}.
\end{multline}
Expansion of the binomials exposes a $k!$ in the numerator, but is also multiplied
by a factor for all possible choices of visiting previously used nodes, leading to a combinatorial
number qualitatively similar to $A_1$. One can rewrite the last expression slightly more compactly as
\begin{multline}
Q_{r-1}\left(k,\left\lceil\frac{k}{r-1}\right\rceil \right)
=\frac{{k\choose r-1}}{\left(\left\lceil\frac{k}{r-1}\right\rceil \right)!}
\sum_{d=0}^{r-1-g}{r-1\choose g+d}\\
\sum_{[d_2+\dots+d_j=d]}
{k-g-d\choose r-1,r-1-d_2,\dots,r-1-d_j}
\prod_{t=1}^{\left\lfloor\frac{k}{r-1}\right\rfloor} {(t-1)(r-1)-\sum_{t'=2}^{t-1}d_{t'}\choose d_t},
\end{multline}
where the multinomial notation 
\begin{equation}
{k-g-d\choose r-1,r-1-d_2,\dots,r-1-d_j}=\frac{[k-g-d]!}{(r-1)!(r-1-d_2)!\dots (r-1-d_j)!}
\end{equation}
has been used.
\section{Discussion and Conclusions}
\label{conclusions}
In this article, I calculate the node neighbor ensemble distribution for random homogeneous $r$-uniform
hypergraphs, or the equivalent problem of the degree distribution in graph ensembles that originate as one-mode
projections of such hypergraph ensembles, giving a precise characterization of the number of unique node
neighbors that a given node possesses on these models.
The relevant qualitative feature of this study is that node overlaps are properly accounted for, so that
no overcounting of neighbors occurs in the distribution.
The sparse and dense limit asymptotics of $\psi_i(k_i,p)$ are also presented. 
These asymptotics provide a way to determine the errors made by ignoring overlaps when computing 
$\psi_i(k_i,p)$, which prove to be asymptotically small in the sparse limit,
but fully dominant in the dense limit. To perform the calculation of the neighbor distribution, 
the quantity $Q_{r-1}$ is introduced and studied for the first time, and its exact formula is provided.
It is worth mentioning that the assembly procedure to calculate $Q_{r-1}$ can be generalized to address
the full problem of mixed rank hypergraphs (or bipartite networks) by extending the
combinatoric rules presented here in Sec.~\ref{Qr-extension}, Eq.~(\ref{xi-r-rules}) to multiple $r$.
It is the author's believe that this work will prove useful in the analysis of theoretical and empirical
problems of systems in which multiway interactions play a dominant role, thus requiring hypergraph or
bipartite network representations.

The author thanks O. Riordan and A. Gerig for helpful discussions
and TSB/EPSRC grant SATURN (TS/H001832/1), ICT eCollective EU project (238597),
and the James Martin 21st Century Foundation Reference no: LC1213-006
for financial support.

\clearpage
\begin{sidewaystable}
\begin{tabular}{|c|c|c||c|c|c|c|c|c|c|c|c|c|c|c|c|}
\hline
$k$&$\ell_{\rm min}$&$\ell_{\rm max}$&$Q_2(k,\ell_{\rm min})$&$Q_2(,+1)$&
$Q_2(,+2)$&$Q_2(,+3)$&$Q_2(,+4)$&$Q_2(,+5)$&$Q_2(,+6)$&$Q_2(,+7)$&
$Q_2(,+8)$&$Q_2(,+9)$&$Q_2(,+10)$&$Q_2(,+11)$&$Q_2(,+12)$\\\hline\hline
2&1&1&1&-&-&-&-&-&-&-&-&-&-&-&-\\\hline
3&2&3&3&1&-&-&-&-&-&-&-&-&-&-&-\\\hline
4&2&6&3&16&15&6&1&-&-&-&-&-&-&-&-\\\hline
5&3&10&30&135&222&205&120&45&10&1&-&-&-&-&-\\\hline
6&3&15&15&330&1581&3760&5715&6165&4945&2997&1365&455&105&15&1\\\hline
\end{tabular}
\caption{A few values of $Q_2(k,\ell)$ with $\ell_{\rm min}=\left\lceil\frac{k}{2} \right\rceil$
and $\ell_{\rm max}={k\choose 2}$. The notation $Q_2(,+x)\equiv Q_2(k,\ell_{\rm min}+x)$.}
\label{Q2table}
\end{sidewaystable}
\clearpage
%
% FIGURE MAP
% fig1-v2.eps fig1-P-illustration.eps
% fig2-v2.eps fig-ki-illustration-v2.eps
% fig3a-v2.eps pk-hk-n32-p0.02-0.05-rea10_5-v3.eps
% fig3b-v2.eps pk-hk-n32-r4-k10-20_rea10_4.eps
% fig4a-v2.eps pk_n128_k4_r3_sparse_sim_bands_theory.eps
% fig4b-v2.eps pk_n64_k6_r4_theory_dsct_sim.eps
% fig4c-v2.eps pk_n2048_k3.5_r4_sparse_bands_theory.eps
% fig4d-v2.eps pk_n64_r3_p0.1-0.2_dense_theory.eps
% fig5a-v2.eps ksparse_k_ratio_N100_r3-4.eps
% fig5b-v2.eps kl_ratio_n100_r3-4_l1-20.eps
% fig6-v2.eps assembly-illustration-v2.eps
% fig7a-v2.eps logQ2-v4.eps
% fig7b-v2.eps logQ2overBinom-v3.eps
%
\clearpage
\clearpage
\clearpage

\begin{thebibliography}{99}
\bibitem{fn-complement} This article serves as a complement to~\cite{Lopez-projections} in which
a more complete study of projections of hypergraphs onto networks is undertaken, but only the
most basic facts are provided about the distribution of degree of the projected graphs.
\bibitem{fn-I} One additional definition of $Q_{r-1}(k_i,\ell_i)$ can be given, visibly relevant in our context: 
$Q_{r-1}(k_i,\ell_i)$ is the cardinality of the set of $\ell_i$ hyperedges of rank $r$ connected to $i$ that 
visit exactly $k_i$ nodes. In Ref.~\cite{Lopez-projections}, such sets where named $\boldsymbol{I}_i(\rho(k_i),\ell_i)$.
\end{thebibliography}
\end{document}